\documentclass{amsart}

\usepackage{amssymb,amsmath,amsthm}
\usepackage{mathrsfs}         

\usepackage{graphicx}        
\usepackage{hyperref}

\usepackage{a4wide}

\theoremstyle{plain}
\newtheorem{thm}{Theorem}
\newtheorem{prop}[thm]{Proposition}

\theoremstyle{definition}
\newtheorem{definition}[thm]{Definition}
\newtheorem{remark}[thm]{Remark}

\newcommand{\tn}[1]{\ensuremath{\mathbb{T}^{#1}}}

\newcommand{\rn}[1]{\ensuremath{\mathbb{R}^{#1}}}
\newcommand{\nn}[1]{\ensuremath{\mathbb{N}^{#1}}}

\newcommand{\sn}[1]{\ensuremath{\mathbb{S}^{#1}}}

\newcommand{\g}{\gamma}
\renewcommand{\a}{\alpha}

\newcommand{\mfg}{\mathfrak{g}}
\newcommand{\e}{\epsilon}
\newcommand{\bK}{\bar{K}}
\newcommand{\bn}{\bar{n}}

\newcommand{\bGa}{\bar{\Gamma}}

\newcommand{\bS}{\bar{S}}
\newcommand{\bge}{\bar{g}}
\newcommand{\bk}{\bar{k}}
\newcommand{\bM}{\bar{M}}
\newcommand{\bx}{\bar{x}}

\newcommand{\robas}{\mathrm{bas}}
\newcommand{\rodiv}{\mathrm{div}}
\newcommand{\rograd}{\mathrm{grad}}
\newcommand{\rotr}{\mathrm{tr}}

\newcommand{\roin}{\mathrm{in}}
\newcommand{\roout}{\mathrm{out}}

\newcommand{\bh}{\bar{h}}

\newcommand{\bD}{\bar{D}}
\newcommand{\bnabla}{\overline{\nabla}}

\newcommand{\hU}{\hat{U}}
\newcommand{\hN}{\hat{N}}

\newcommand{\chh}{\check{h}}

\newcommand{\mK}{\mathcal{K}}

\newcommand{\msK}{\mathscr{K}}
\newcommand{\msI}{\mathscr{I}}
\newcommand{\msG}{\mathscr{G}}
\newcommand{\msO}{\mathscr{O}}
\newcommand{\msX}{\mathscr{X}}

\newcommand{\msY}{\mathscr{Y}}

\newcommand{\tr}{\mathrm{tr}}
\renewcommand{\d}{\partial}

\begin{document}

\author{Hans Ringstr\"{o}m}
\title{Initial data on big bang singularities in symmetric settings}
\begin{abstract}
  In a recent article, we propose a general geometric notion of initial data on big bang singularities. This notion is of interest in its own
  right. However, it also serves the purpose of giving a unified perspective on many of the results in the literature. In the present article,
  we give a partial justification of this statement by rephrasing the results concerning Bianchi class A orthogonal stiff solutions
  and solutions in the $\tn{3}$-Gowdy symmetric vacuum setting in terms of our general geometric notion of initial data on the big bang singularity.
\end{abstract}

\maketitle

\section{Initial data on the big bang singularity}

In the last 20--30 years, a substantial number of results concerning quiescent big bang singularities have appeared; cf., e.g.,
\cite{ptc,ciam,kar,iak,ren,aarendall,raw,sta,damouretal,iam,asvelGowdy,SCCGowdy,ABIF,klinger,aeta,rasq,rsh,specks3,fal,GIJ,amesetal,amesetalLambda,bao}.
In some of the results, solutions are constructed given initial data on the singularity. However, how the data are specified often depends on
the particular choice of foliation or on the specific symmetry class under consideration. The purpose of the present article is to take the first
step towards illustrating that the quiescent solutions studied in the literature can be thought of as arising from solutions to a set of general
geometric conditions, including constraint equations, on the big bang singularity. In the context of the Einstein-scalar field equations or in the
case of Einstein's vacuum equations in higher dimensions ($d+1$, where $d\geq 10$), the constraint equations are sufficient; cf.
\cite{aarendall,damouretal} and \cite[Subsection~1.5]{RinQC}. However, in the
$3+1$-dimensional vacuum setting, e.g., an additional condition has to be satisfied in order for quiescent solutions to exist. We begin by
formulating this quiescence criterion in terms of the data at the singularity; cf. \cite{RinQC}.


\subsection{Quiescent initial data on the singularity} In the case of the Einstein-scalar field equations with a cosmological constant $\Lambda$, 
quiescent initial data on the singularity take the form $(\bM,\bh,\msK,\Phi_{a},\Phi_{b})$. Here $\bM$ is a $3$-dimensional manifold, $\bh$ is a
Riemannian metric on $\bM$, $\msK$ is a $(1,1)$-tensor field on $\bM$ and $\Phi_{a}$ and $\Phi_{b}$ are scalar functions on $\bM$. We think of $\msK$
as an endomorphism of the tangent bundle of $\bM$. Assume the eigenvalues of $\msK$ to be real and distinct, denote them by $p_{A}$ and assume them to be
ordered so that $p_{1}<p_{2}<p_{3}$. Let $\{\msX_{A}\}$ be a corresponding local frame of eigenvector fields of $\msK$; i.e., $\msK\msX_{A}=p_{A}\msX_{A}$
(no summation). Assume, moreover, $\{\msX_{A}\}$ to be unit vector fields with respect to $\bh$. These conditions define the local vector fields
$\msX_{A}$ up to a sign. This means that the functions $\g^{A}_{BC}$, given by $[\msX_{B},\msX_{C}]=\g^{A}_{BC}\msX_{A}$, are well defined up to a sign.
Introducing $\msO^{A}_{BC}:=(\g^{A}_{BC})^{2}$, it is thus clear that $\msO^{A}_{BC}$ are globally well defined smooth functions on $\bM$. 

\begin{definition}\label{def:ndvacidonbbs}
  Let $\Lambda\in\rn{}$, $(\bM,\bh)$ be a smooth $3$-dimensional Riemannian manifold, $\msK$ be a smooth $(1,1)$-tensor field on
  $\bM$ and $\Phi_{a}$ and $\Phi_{b}$ be smooth functions on $\bM$. Then $(\bM,\bh,\msK,\Phi_{a},\Phi_{b})$ are
  \textit{non-degenerate quiescent initial data on the singularity for the Einstein-scalar field equations with a cosmological constant $\Lambda$}
  if, with notation as above, 
  \begin{enumerate}
  \item $\tr\msK=1$ and $\msK$ is symmetric with respect to $\bh$.
  \item $\tr\msK^{2}+\Phi_{a}^{2}=1$ and $\mathrm{div}_{\bh}\msK=\Phi_{a}d\Phi_{b}$.
  \item The eigenvalues of $\msK$ are distinct.
  \item $\msO^{k}_{ij}$ vanishes in a neighbourhood of $\bx\in\bM$ if $p_{k}(\bx)\leq 0$ and $\{i,j,k\}=\{1,2,3\}$.    
  \end{enumerate}
\end{definition}
\begin{remark}
  That $\msK$ is symmetric with respect to $\bh$ means that if $p\in\bM$ and $\xi,\zeta\in T_{p}\bM$, then $\bh(\msK\xi,\zeta)=\bh(\xi,\msK\zeta)$.
  This condition implies that the eigenvalues of $\msK$ are real. 
\end{remark}
\begin{remark}
  The definition is a special case of \cite[Definition~10]{RinQC} which holds for spacetime dimensions $d+1$, $d\geq 3$. 
\end{remark}

\subsection{Developments corresponding to data on the singularity}\label{ssection:developments}
Given non-degenerate quiescent initial data on the singularity as in Definition~\ref{def:ndvacidonbbs}, the goal is to demonstrate that there is a unique
corresponding maximal globally hyperbolic development, say $(M,g,\phi)$. Exactly how the data on the singularity correspond to a solution depends
on the choice of foliation of the relevant spacetime. Here we focus on the case of foliations with a vanishing shift vector field and uniformly
diverging mean curvature, though constructing, e.g., Gaussian foliations is also a possibility (cf. \cite{RinQC} for a discussion of this case).
The goal is to prove that there is a unique maximal
globally hyperbolic development, say $(M,g,\phi)$, solving the Einstein-scalar field equations with a cosmological constant $\Lambda$ such that the
following holds. There is a $0<t_{+}\in\rn{}$ and a diffeomorphism $\Psi$ from $\bM\times (0,t_{+})$ to an open subset of $(M,g)$ such that 
$\Psi^{*}g$ can locally be represented as
\begin{equation}\label{eq:rinansatzintrocrushing}
  (\Psi^{*}g)=-N^{2}dt\otimes dt+\textstyle{\sum}_{A,B}b_{AB}\theta^{-2p_{\max\{A,B\}}}\msY^{A}\otimes \msY^{B},
\end{equation}
where $N$ is a strictly positive function (the \textit{lapse function}), $\theta$ is the mean curvature of the leaves of the foliation, $\{\msX_{A}\}$
is a local basis of eigenvector fields of $\msK$ and $\{\msY^{A}\}$ is the dual basis. Moreover, $\msX_{A}$ corresponds to the eigenvalue $p_{A}$ and
the $p_{A}$ are ordered so that $p_{1}<p_{2}<p_{3}$. Here, we require $\theta$ to diverge to infinity uniformly as $t\rightarrow 0+$. Let $\bK$ be the
Weingarten map of the leaves of the foliation (i.e., the second fundamental form with one index raised by the induced metric); $\mK:=\bK/\theta$ be
the \textit{expansion normalised Weingarten map}; and $\chh$ be defined by
\[
\chh:=\textstyle{\sum}_{A,B}b_{AB}\msY^{A}\otimes \msY^{B};
\]
note that $\chh$ is globally well defined independent of the choice of local frame. Then we require the following correspondence between
the solution and the data on the singularity
\begin{equation}\label{eq:mKlimetc}
  \lim_{t\rightarrow 0+}\mK = \msK,\ \ \
  \lim_{t\rightarrow 0+}\chh = \bh,\ \ \
  \lim_{t\rightarrow 0+}(\hU\phi) = \Phi_{a},\ \ \
  \lim_{t\rightarrow 0+}(\phi+\Phi_{a}\ln \theta) = \Phi_{b},
\end{equation}
where $\hU:=\hN^{-1}\d_{t}$ and $\hN:=\theta N$. In practice, the exact function space in which this convergence takes place should be specified.
Moreover, in certain circumstances it might be natural to require a particular rate of convergence; cf. \cite{RinQC} for a more detailed discussion.
It might also be necessary to impose additional conditions on the foliation in order to ensure existence and uniqueness. 

\subsection{Comments on the conditions}
It is of interest to comment on the conditions appearing in Definition~\ref{def:ndvacidonbbs}. We do so in the form of the following remarks. 
\begin{remark}
  The condition that $\tr\msK=1$ arises from (\ref{eq:mKlimetc}) and the fact that $\tr\mK=1$ by definition. The conditions
  $\mathrm{div}_{\bh}\msK=\Phi_{a}d\Phi_{b}$ and $\tr\msK^{2}+\Phi_{a}^{2}=1$ are what remains of the momentum and Hamiltonian constraints
  respectively.
\end{remark}
\begin{remark}
  In terms of $\Phi_{a}$ and the $p_{A}$, the conditions $\tr\msK=1$ and $\tr\msK^{2}+\Phi_{a}^{2}=1$ translate to
  \begin{equation}\label{eq:KasnerRelations}
    \textstyle{\sum}_{A}p_{A}=1,\ \ \
    \textstyle{\sum}_{A}p_{A}^{2}+\Phi_{a}^{2}=1
  \end{equation}
  respectively. In vacuum, these conditions correspond to the requirement that the eigenvalues of $\msK$ satisfy the Kasner relations. 
\end{remark}
\begin{remark}
  Due to the framework developed in \cite{RinGeo}, we expect the last criterion in Definition~\ref{def:ndvacidonbbs} to be necessary in order to
  obtain quiescence. However, if $p_{1}>0$, this condition is void. 
\end{remark}
\begin{remark}
  In general, there is not a global frame of eigenvector fields of $\msK$. However, by going to a finite covering space (of $\bM$), if necessary,
  the frame can be assumed to be global; cf. \cite[Lemma~A.1, p.~201]{RinWave}. In that setting, the equality (\ref{eq:rinansatzintrocrushing})
  holds globally.
\end{remark}
\begin{remark}
  The requirement of non-degeneracy is unfortunate. However, in the case of Gaussian foliations, it might be possible to avoid it;
  cf. \cite[Remark~26]{RinQC}.
\end{remark}

\subsection{Spatially homogeneous setting}
It is of interest to interpret the results in the spatially homogeneous setting in the light of the above notion of initial data on the singularity. 

\textit{Regular initial data.} In the $3+1$-dimensional spatially homogeneous setting, there are three cases to consider: left
invariant initial data on a unimodular Lie group (Bianchi class A); left invariant initial data on a non-unimodular Lie group (Bianchi class B); and
initial data on $\sn{2}\times\rn{}$ invariant under the isometry group of the standard Riemannian metric on $\sn{2}\times\rn{}$ (Kantowski-Sachs).
Here we focus solely on Bianchi class A with orthogonal perfect fluid matter. To be more precise, the corresponding initial data are defined as follows
(cf. \cite[Definition~9, p.~607]{KGCos}):
\begin{definition}\label{def:Bianchiid}
  \textit{Bianchi class A orthogonal perfect fluid initial data} for Einstein's equations consist of the following: a connected $3$-dimensional
  unimodular Lie group $G$; a left invariant metric $\bge$ on $G$; a left invariant symmetric covariant $2$-tensor field $\bk$ on $G$; and a constant
  $\rho_{0}\geq 0$ satisfying
  \begin{align*}
    \bS-\bk^{ij}\bk_{ij}+(\rotr_{\bge}\bk)^{2} = & 2\rho_{0}\\
    \bnabla_{i}\rotr_{\bge}\bk-\bnabla^{j}\bk_{ij} = & 0.
  \end{align*}
\end{definition}
\textit{Developments.} Given Bianchi class A orthogonal perfect fluid initial data for Einstein's equations and a constant $\g\in (2/3,2]$, there
is a corresponding Bianchi class A development with equation of state $p=(\g-1)\rho$; cf. \cite[Definition~21.1, p. 489]{BianchiIXattr}. The metric
takes the form
\begin{equation}\label{eq:gdiagonalSH}
  g=-dt\otimes dt+\textstyle{\sum}_{i}a_{i}^{2}(t)\xi^{i}\otimes \xi^{j}
\end{equation}
on $G\times I$, where $I$ is an open interval. Moreover, the stress energy tensor of the orthogonal perfect fluid takes the form
\[
T:=(\rho+p)dt\otimes dt+pg.
\]
In the above expressions, $\rho$ denotes the \textit{energy density} and $p$ the \textit{pressure};
$a_{i}$ are smooth positive functions of $t$; $\{e_{i}\}$ is a basis of the Lie algebra
$\mathfrak{g}$ of $G$; $\{\xi^{i}\}$ is the dual basis of $\{e_{i}\}$; and there are constants $n_{k}$ such that
\begin{equation}\label{eq:diagonaln}
  [e_{i},e_{j}]=\e_{ijk}n_{k}e_{k}
\end{equation}
(no summation on $k$). The constants $n_{k}$ correspond to a classification of the Lie algebras: all $n_{i}=0$ (Bianchi type I); all but one of the
$n_{i}=0$ (Bianchi type II); two $n_{i}$ non-zero, different signs (Bianchi type VI${}_{0}$); two $n_{i}$ non-zero, same signs (Bianchi type
VII${}_{0}$); all $n_{i}$ non-zero, not all same sign (Bianchi type VIII); all $n_{i}$ non-zero, all have the same sign (Bianchi type IX).

The above developments are globally hyperbolic; cf. \cite[Lemma~21.4, p.~490]{BianchiIXattr}. Moreover, in
case $\rho_{0}$ appearing in Definition~\ref{def:Bianchiid} is non-zero, the spacetime Ricci tensor, contracted with itself, diverges in
directions that are not geodesically complete; cf. \cite[Lemma~22.3, p.~497]{BianchiIXattr}. This means that the developments are
$C^{2}$-inextendible if $\rho_{0}\neq 0$. In particular, they are the maximal globally hyperbolic developments of the initial data. In case
$\rho_{0}=0$, then, for a given time direction, either the mean curvature diverges or the spacetime is geodesically complete. This means that
the corresponding developments are inextendible as globally hyperbolic developments; cf. \cite[Proposition~18.16, p.~203]{RinCauchy}. For a complete
picture in the vacuum setting, see \cite[Theorem~24.12, p.~258]{RinCauchy}.

\textit{Correspondence between orthogonal stiff fluids and massless scalar fields in the spatially homogeneous setting.} The orthogonal perfect
fluids considered above include \textit{dust} ($\g=1$), \textit{radiation} ($\g=4/3$) and \textit{stiff fluids} ($\g=2$). In what follows, it is of
interest to keep in mind that a stiff fluid can be interpreted as a massless scalar field and vice versa. To be more precise, let $\phi$ be a massless
scalar field on $(M,g)$, where $g$ is of the form (\ref{eq:gdiagonalSH}) and $M:=G\times I$, and assume that $\phi$ only depends on $t$. Then,
defining $\rho=\dot{\phi}^{2}/2$, the stress energy tensor associated with $\phi$ can be interpreted as the stress energy tensor of an orthogonal
stiff fluid. Alternately, given an orthogonal stiff fluid solution with non-vanishing energy density $\rho$ (note that if $\rho=0$ at one point in time,
$\rho=0$ for all $t$), we can define a scalar field by integrating the relation $\phi_{t}=(2\rho)^{1/2}$. 

\textit{Data on the singularity in the Bianchi class A setting.} Due to the above equivalence between massless scalar fields and orthogonal stiff fluids,
and since the results to which we appeal are stated in the stiff fluid setting, we use the following formulation of data on the singularity
in the Bianchi class A setting with a vanishing cosmological constant.

\begin{definition}\label{def:ndvacidonbbssh}
  Let $G$ be a $3$-dimensional unimodular Lie group, $\bh$ be a left invariant Riemannian metric on $G$, $\msK$ be a left invariant
  $(1,1)$-tensor field on $G$ and $\omega$ be a non-negative real number. Then $(G,\bh,\msK,\omega)$ are
  \textit{non-degenerate quiescent Bianchi class A initial data on the singularity for the Einstein-orthogonal stiff fluid equations} if
  \begin{enumerate}
  \item $\tr\msK=1$ and $\msK$ is symmetric with respect to $\bh$,
  \item $\tr\msK^{2}+2\omega/3=1$ and $\mathrm{div}_{\bh}\msK=0$,
  \item the eigenvalues of $\msK$ are distinct,
  \item $\msO^{k}_{ij}=0$ if $p_{k}\leq 0$, where $\{i,j,k\}=\{1,2,3\}$. 
  \end{enumerate}
\end{definition}
\begin{remark}
  Here $p_{A}$ are the eigenvalues of $\msK$, and they are constants due to the left invariance of $\msK$. 
  In case $\omega=0$, we speak of non-degenerate quiescent Bianchi class A initial data on the singularity for the Einstein vacuum equations.
\end{remark}
Before discussing the existence of initial data on the singularity, as well as the existence and uniqueness of corresponding developments, it is
useful to rephrase the conditions. To begin with, due to \cite[Corollary~19.14, p.~211]{RinCauchy}, the conditions that $\msK$ is symmetric with
respect to $\bh$ and that $\mathrm{div}_{\bh}\msK=0$ are equivalent to the existence of a basis $\{e_{i}\}$ of the Lie algebra $\mathfrak{g}$ such that
\begin{equation}\label{eq:canonicalbasisBianchiA}
  \bh(e_{i},e_{j})=\delta_{ij},\ \ \
  \msK e_{i}=p_{i}e_{i},\ \ \
  [e_{i},e_{j}]=\e_{ijk}n_{k}e_{k},
\end{equation}
where there is no summation on $i$ in the second equality, no summation on $k$ in the third equality, $\e_{123}=1$ and $\e_{ijk}$ is antisymmetric
under permutations of the indices. Here $p_{i}$ and $n_{k}$ are constants. Note that the $p_{i}$ are the eigenvalues of $\msK$. Introducing
$p_{\pm}$ by
\[
p_{+}:=\frac{3}{2}\left(p_{2}+p_{3}-\frac{2}{3}\right),\ \ \
p_{-}:=\frac{\sqrt{3}}{2}(p_{2}-p_{3}),
\]
then the relation $\tr\msK=1$ is automatically satisfied and the relation $\tr\msK^{2}+2\omega/3=1$ can be rephrased
\[
\omega+p_{+}^{2}+p_{-}^{2}=1.
\]
Finally, the fourth condition can be rephrased as saying that $n_{k}=0$ if $p_{k}\leq 0$. Note that one particular consequence of the last
observation is that there are only initial data on the singularity of Bianchi types VIII and IX in case all the eigenvalues of $\msK$ are
strictly positive. Since this is not possible for $\omega=0$, we conclude that for Bianchi types VIII and IX, there are no non-degenerate
quiescent Bianchi class A initial data on the singularity for the Einstein vacuum equations. This corresponds to the fact that all Bianchi
type VIII and IX vacuum solutions that are not locally rotationally symmetric have oscillatory singularities; cf. \cite{cbu} for details.
\begin{figure}
  \begin{center}
    \includegraphics{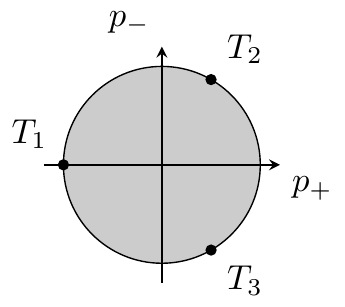}
    \includegraphics{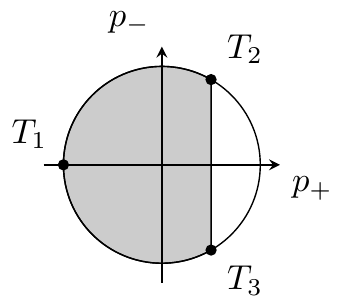}
    \includegraphics{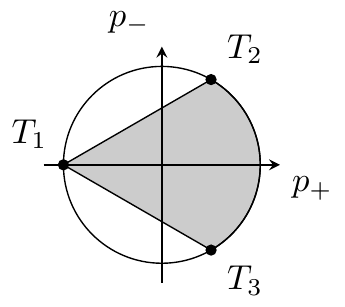}
    \includegraphics{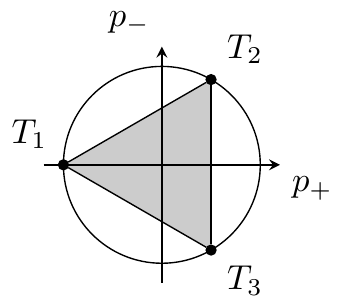}
  \end{center}
  \caption{The gray areas indicate the regimes of $p_{+}$, $p_{-}$ for which there are corresponding non-degenerate quiescent Bianchi class A
    initial data on the singularity for the Einstein-orthogonal stiff fluid equations. The first image from the left corresponds to Bianchi type I
    (all $n_{k}$ vanish); the second image corresponds to Bianchi type II (with $n_{1}\neq 0$); the third image corresponds to Bianchi types VI${}_{0}$
    and VII${}_{0}$ (with $n_{2}\neq 0$ and $n_{3}\neq 0$); and the third image corresponds to Bianchi types VIII and IX (with all $n_{i}$ non-vanishing).
    Note that the admissible non-degenerate data are obtained by removing the line segments connecting each of the $T_{i}$ with its antipodal
    point. Moreover, the vacuum setting corresponds to the admissible points on the unit circle.}  
  \label{fig:QuiescentConvergentRegime}
\end{figure}
The admissible subsets of the $p_{+}p_{-}$-plane corresponding to data on the singularity is illustrated in Figure~\ref{fig:QuiescentConvergentRegime}.

\textit{Results.} Next, we state the results concerning the asymptotics for vacuum and orthogonal stiff fluid solutions.

\begin{thm}\label{thm:convergence}
  Consider Bianchi class A orthogonal stiff fluid initial data and let $(M,g,\rho)$ be the corresponding orthogonal stiff fluid development, where
  $M:=G\times I$,
  $g$ takes the form (\ref{eq:gdiagonalSH}) and $\rho$ is the energy density. In the vacuum setting, assume that the development is neither of Bianchi
  type VIII nor IX; that it is not a quotient of Minkowski space; and that it does not admit an extension through a Cauchy horizon. Then, given an
  appropriate time orientation, the mean curvature of the hypersurfaces of spatial homogeneity, say $\theta$, decreases strictly from $\infty$ to
  $0$ in case the development is not of type IX, and decreases strictly from $\infty$ to $-\infty$ in case the development is of type IX. Moreover,
  the existence interval $I$ is finite to the past and can be assumed to take the form $I=(0,t_{+})$, with $t_{+}\leq \infty$. For $t$ close to $0$,
  $\theta>0$ and $\Omega=3\rho/\theta^{2}$ and the expansion normalised Weingarten map associated with $g$, say $\mK$, are well defined.

  Under the above assumptions, $\Omega$ and $\mK$ converge to limits, say $\omega$ and $\msK$ respectively, as $t\rightarrow 0+$. These limits are such
  that $\tr\msK=1$ and $\tr\msK^{2}+2\omega/3=1$. Moreover, there is a choice of basis $\{e_{i}\}$ of $\mathfrak{g}$ such that the $e_{i}$ are orthogonal
  with respect to $g$ (so that the $\xi^{i}$ appearing in (\ref{eq:gdiagonalSH}) can be assumed to belong to the basis dual to $\{e_{i}\}$); such
  that (\ref{eq:diagonaln}) holds; such that $\msK$ is diagonal with respect to $\{e_{i}\}$; and such that
  \begin{equation}\label{eq:tmpiailimit}
    \lim_{t\rightarrow 0+}t^{-p_{i}}a_{i}(t)=1,\ \ \
    \lim_{t\rightarrow 0+}\theta^{p_{i}}a_{i}(t)=1,
  \end{equation}
  where $p_{i}$ is the eigenvalue of $\msK$ corresponding to $e_{i}$. In particular,
  \begin{equation}\label{eq:gasymptoticformthm}
    g=-dt\otimes dt+\sum b_{i}^{2}t^{2p_{i}}\xi^{i}\otimes\xi^{i},\ \ \
    g=-dt\otimes dt+\sum \bar{b}_{i}^{2}\theta^{-2p_{i}}\xi^{i}\otimes\xi^{i},
  \end{equation}
  where
  \begin{equation}\label{eq:limitofchhsphom}
    \lim_{t\rightarrow 0+}\sum b_{i}^{2}\xi^{i}\otimes\xi^{i}
    =\lim_{t\rightarrow 0+}\sum\bar{b}_{i}^{2}\xi^{i}\otimes\xi^{i}=\sum\xi^{i}\otimes\xi^{i}=:\bh.
  \end{equation}
  Finally, $\msK$ is symmetric with respect to $\bh$; $\rodiv_{\bh}\msK=0$; and $\msO^{k}_{ij}=0$ if $p_{k}\leq 0$, where $\{i,j,k\}=\{1,2,3\}$. 
\end{thm}
\begin{remark}\label{remark:quantitativelimits}
  The conclusions can be improved in the sense that there are $t_{+}>0$, $\epsilon>0$ and $C>0$ such that
  \[
  |\Omega-\omega|+|\theta^{p_{i}}a_{i}-1|+|t\theta-1|+|\mK-\msK|_{\bh}+|\chh-\bh|_{\bh}\leq C\theta^{-\epsilon}
  \]
  on $(0,t_{+})$, where $\chh:=\sum \bar{b}_{i}^{2}\xi^{i}\otimes\xi^{i}$.
\end{remark}
\begin{remark}
  The specific assumptions concerning the initial data that give rise to quotients of Minkowski space and Cauchy horizons are
  described in \cite[Theorem~24.12, p.~258]{RinCauchy}.
\end{remark}
\begin{remark}
  Under the assumptions of the theorem, we obtain data on the singularity as in Definition~\ref{def:ndvacidonbbssh}, with the exception of the
  condition of non-degeneracy, which we do not address here.
\end{remark}
\begin{proof}
  The proof is to be found in Subsection~\ref{ssection:asymptoticshs}.
\end{proof}

It is of interest to ask the reverse question: Given data as in Definition~\ref{def:ndvacidonbbssh}, is there a unique corresponding
development?

\begin{thm}\label{thm:dataonsingtosolution}
  Let $(G,\bh,\msK,\omega)$ be non-degenerate quiescent Bianchi class A initial data on the singularity for the Einstein-orthogonal stiff fluid equations.
  Then there is a unique Bianchi class A stiff fluid development such that $\theta\rightarrow\infty$ as $t\rightarrow t_{-}$ (where $t_{-}$ is the
  left endpoint of the existence interval), and such that if $\mK$ is the expansion normalised Weingarten map of the development
  and $\Omega=3\rho/\theta^{2}$, where $\rho$ is the energy density, then the following holds:
  \begin{enumerate}
  \item $\mK\rightarrow\msK$ and $\Omega\rightarrow\omega$ as $t\rightarrow t_{-}$. 
  \item If $\{e_{i}\}$ is chosen to be a basis for $\mathfrak{g}$ consisting of eigenvectors of $\msK$, where $\msK e_{i}=p_{i}e_{i}$ (no summation)
    and $p_{1}<p_{2}<p_{3}$, then the metric $g$ of the development can be written
    \begin{equation}\label{eq:sphomasformcrush}
      g=-dt\otimes dt+\textstyle{\sum}_{i}b_{i}^{2}\theta^{-2p_{i}}\xi^{i}\otimes\xi^{j}
    \end{equation}
    for $t\leq t_{0}$ and some $t_{0}>t_{-}$ belonging to the existence interval, where $\{\xi^{i}\}$ is the basis dual to $\{e_{i}\}$.
  \item The following limit holds:
    \[
    \lim_{t\rightarrow t_{-}}\textstyle{\sum}_{i}b_{i}^{2}\xi^{i}\otimes\xi^{j}=\bh.
    \]
  \end{enumerate}  
\end{thm}
\begin{remark}
  A similar statement holds with the form (\ref{eq:sphomasformcrush}) replaced by
  \[
  g=-dt\otimes dt+\textstyle{\sum}_{i}b_{i}^{2}t^{2p_{i}}\xi^{i}\otimes\xi^{j}.
  \]
  In other words, we can think of the foliation of the Bianchi class A development as a CMC foliation or as a Gaussian foliation. Existence and
  uniqueness holds irrespective of perspective. 
\end{remark}
\begin{remark}
  In case $\omega=0$, the relevant developments are vacuum developments.
\end{remark}
\begin{remark}
  The uniqueness statement holds under somewhat weaker assumptions; we do not need to assume the $\{\xi^{i}\}$ appearing in
  (\ref{eq:sphomasformcrush}) to be the duals of an eigenframe of $\msK$ etc. In fact, the uniqueness holds in the class of Bianchi class
  A developments; cf. the proof. 
\end{remark}
\begin{proof}
  The proof is to be found in Subsection~\ref{ssection:proofofgeometricunique}.
\end{proof}

\subsection{$\tn{3}$-Gowdy symmetric vacuum spacetimes}
As a first step beyond the spatially homogeneous setting, it is natural to consider $\tn{3}$-Gowdy symmetric vacuum spacetimes. For the purposes of the
present discussion, $\tn{3}$-Gowdy metrics are taken to be of the form
\begin{equation}\label{eq:Gowdymetricareal}
  g=t^{-1/2}e^{\lambda/2}(-dt^{2}+d\vartheta^{2})+te^{P}(dx+Qdy)^{2}+te^{-P}dy^{2},
\end{equation}
where $\lambda$, $P$ and $Q$ only depend on $t$ and $\vartheta$; we refer the interested reader to \cite{gowdy,ptc} for a description of the origin
of this class of spacetimes. The underlying manifold is
$M:=\tn{3}\times (0,\infty)$. However, from a geometric perspective, it is natural to view it as $\sn{1}\times\tn{2}\times (0,\infty)$. Moreover, the
metric is invariant under translations in the $\tn{2}$-factor, $t$ is
the coordinate on the $(0,\infty)$-factor and $\vartheta$ is the ``coordinate'' on the $\sn{1}$-factor. The symbols $x$ and $y$ can be thought
of as coordinates on the universal covering space of $\tn{2}$; i.e., on $\rn{2}$. These coordinates do, of course, not descend to the quotient,
but the frame $\{dx,dy\}$ of the cotangent space does. Here it should be emphasised that $\d_{x}$ and $\d_{y}$ do not necessarily correspond to the
two $\sn{1}$-factors in $\tn{2}$. The time coordinate $t$ in (\ref{eq:Gowdymetricareal}) is such that the area of the symmetry orbits (i.e., the
area of the $\tn{2}$'s) is proportional to $t$. Initial data for a metric of the form (\ref{eq:Gowdymetricareal}) are given by $\lambda$, $P$ and
$Q$ and their time derivatives. However, the Hamiltonian and momentum constraints translate to
\begin{subequations}\label{eq:lambda}
  \begin{align}
    \lambda_{t} = & t[P_{t}^{2}+P_{\vartheta}^{2}+e^{2P}(Q_{t}^{2}+Q_{\vartheta}^{2})],\label{eq:lambdat}\\
    \lambda_{\vartheta} = & 2t(P_{t}P_{\vartheta}+e^{2P}Q_{t}Q_{\vartheta});\label{eq:lambdatheta}
  \end{align}
\end{subequations}
cf., e.g., \cite[Section~2]{AAR}. In particular, the initial data for $\lambda$ are, up to a constant, determined by the initial data for $P$
and $Q$. Therefore, the initial data consist of $P$, $Q$,
their time derivatives and a constant (corresponding, say, to the mean value of $\lambda$ at the initial time). Moreover, the only constraint is
that the integral of the right hand side of (\ref{eq:lambdatheta}) over $\sn{1}$ equals zero. For these reasons, the set of initial data, say
$\msI$, consist of four functions in $C^{\infty}(\sn{1})$, say $P(\cdot,t_{0})$, $Q(\cdot,t_{0})$, $P_{t}(\cdot,t_{0})$ and $Q_{t}(\cdot,t_{0})$, such
that the integral of the right hand side of (\ref{eq:lambdatheta}) over $\sn{1}$ equals zero. Effectively, the relevant equations are thus those for
$P$ and $Q$:
\begin{subequations}\label{eq:PQ}
  \begin{align}
    \d_{t}(tP_{t}) = & \d_{\vartheta}(tP_{\vartheta})+te^{2P}(Q_{t}^{2}-Q_{\vartheta}^{2}),\label{eq:Pwave}\\
    \d_{t}(te^{2P}Q_{t}) = & \d_{\vartheta}(te^{2P}Q_{\vartheta});\label{eq:Qwave}
  \end{align}
\end{subequations}
cf., e.g., \cite[Section~2]{AAR}.
This is a system of wave equations for $P$ and $Q$ which can be interpreted as a wave map system with $\mathbb{H}^{2}$ (two dimensional hyperbolic
space) as target; cf., e.g., \cite[p.~980]{asvelGowdy}. It is of interest to compare the asymptotics of metrics of the form
(\ref{eq:Gowdymetricareal}), where $P$, $\lambda$ and $Q$ satsify (\ref{eq:lambda}) and (\ref{eq:PQ}), with Definition~\ref{def:ndvacidonbbs} and
the developments discussed in Subsection~\ref{ssection:developments}. 

\begin{thm}\label{thm:gowdy}
  There is a subset of $\msI$, say $\msG$, which is open and dense in the $C^{k+1}\times C^{k}$-topology for all $1\leq k\leq\infty$, such that the
  following holds. Fix a solution corresponding to an element of $\msG$. Then there is a finite number of points $\vartheta_{i}\in\sn{1}$,
  $i=1,\dots,k$, such that if $S:=\sn{1}-\cup_{i}\{\vartheta_{i}\}$, there is a smooth Riemannian metric $\bh$ and a smooth $(1,1)$-tensor field $\msK$
  on $S\times\tn{2}$ such that $\msK$ is symmetric with respect to $\bh$ and the eigenvalues of $\msK$, say $p_{i}$, $i=1,2,3$, are distinct and such
  that $p_{1}<p_{2}<p_{3}$. Moreover, if $\{\msX_{A}\}$ is a smooth local frame of vectorfields satisfying $\msK \msX_{A}=p_{A}\msX_{A}$ (no summation) and
  $|\msX_{A}|_{\bh}=1$, then
  \[
  g=-N^{2}dt\otimes dt+\textstyle{\sum}_{A,B=1}^{3}c_{AB}\theta^{-2p_{\max\{A,B\}}}\msY^{A}\otimes\msY^{B},
  \]
  where $\{\msY^{A}\}$ is the basis dual to $\{\msX_{A}\}$. In addition, if $J$ is an open interval in $\sn{1}$ with compact closure contained in $S$,
  then there are, for each $l\in\nn{}$, constants $C_{l}$ as well as a constant $\eta>0$ such that
  \begin{subequations}
    \begin{align}
      |\bD^{l}(\chh-\bh)|_{\bge_{0}} \leq & C_{l}\theta^{-\eta},\label{eq:chhconvtobh}\\
      |\bD^{l}(\mK-\msK)|_{\bge_{0}} \leq & C_{l}\theta^{-\eta}\label{eq:mKtomsK}
    \end{align}
  \end{subequations}
  for $\vartheta\in J$ and $t\in (0,1)$, where $\bge_{0}$ is the standard Euclidean metric on $\tn{3}$, $\bD$ is the corresponding Levi-Civita
  connection, $\mK$ is the expansion normalised Weingarten map and
  \[
  \chh:=\textstyle{\sum}_{A,B=1}^{3}c_{AB}\msY^{A}\otimes\msY^{B}.
  \]
  Finally, $\tr\msK=1$; $\tr\msK^{2}=1$; $\rodiv_{\bh}\msK=0$; $\msO^{1}_{23}=0$; and the mean curvature diverges uniformly to $\infty$ in the
  direction of the singularity. 
  
  In addition, $\bh$ and $\msK$ are invariant under the orbits of $\tn{2}$. Moreover, the direction parallel to the $\sn{1}$-factor is perpendicular
  to the symmetry orbits and is an eigendirection of $\msK$. Finally, for $\vartheta\in S$, the $X_{1}=\d_{\vartheta}$ vectorfield is, up to a sign,
  determined by the condition that it is perpendicular to the symmetry orbits and the condition that
  \begin{equation}\label{eq:Xobhform}
    \mathrm{Area}_{\bh_{0}}(\tn{2})=(1-p_{\perp})|X_{1}|_{\bh}\mathrm{Area}_{\bh}(\tn{2}),
  \end{equation}
  where $p_{\perp}$ is the eigenvalue corresponding to $\d_{\vartheta}$; $\bh_{0}:=dx^{2}+dy^{2}$ on $\tn{2}$; and $\{dx,dy\}$ is the frame of one-form
  fields appearing in (\ref{eq:Gowdymetricareal}).

  For all $\vartheta\in\sn{1}$, the eigenvalues of $\mK(\vartheta,t)$ converge as $t\rightarrow 0+$. Moreover, the limits satisfy the Kasner
  relations; i.e., the sum of the limits of the eigenvalues equals $1$, and the sum of the squares of the limits of the eigenvalues equals $1$.
  One of the limits corresponds to the vector field $\d_{\vartheta}$. Denote this limit by $p_{\perp}$. Then
  $p_{\perp}(\vartheta)\in (-1/3,0)$ for all $\vartheta\in S$. Moreover, $p_{\perp}(\vartheta_{i})>0$ for all $i=1,\dots,k$. Finally, denote the
  collection of limits at $\vartheta$ by $p(\vartheta)$. Then $p(\vartheta)$ can be thought of as a point on the Kasner circle. Moreover, the limit
  of $p$ as $\vartheta\rightarrow\vartheta_{i}$ exists. Denote this limit by $p(\vartheta_{i}\pm)$. Then $p(\vartheta_{i}\pm)$ is obtained by applying
  the Kasner map to $p(\vartheta_{i})$.

  Let $\ell_{a}$ and $\ell_{b}$ be the eigenvalues corresponding to eigenvector fields of $\mK$ tangential to the symmetry orbits. Then
  $\ell_{a}(\vartheta_{i},t)$ and $\ell_{b}(\vartheta_{i},t)$ converge to limits, say $p_{a,i}$ and $p_{b,i}$ respectively, as $t\rightarrow 0+$,
  with $p_{a,i}p_{b,i}<0$. Let $\xi_{a}$ and $\xi_{b}$ be eigenvector fields corresponding to $\ell_{a}$ and $\ell_{b}$ respectively, normalised
  with respect to a fixed Riemannian metric on $\tn{3}$. Then $\xi_{a}(\vartheta_{i},t)$ and $\xi_{b}(\vartheta_{i},t)$ converge to limits, say
  $\eta_{a,i}$ and $\eta_{b,i}$ respectively, as $t\rightarrow 0+$. Moreover, $\eta_{a,i}=\pm\eta_{b,i}$. 
\end{thm}
\begin{remark}
  The $\vartheta_{i}$, $i=1,\dots,k$, represent the non-degenerate true spikes of the solution; cf. \cite[Definition~3, p.~1189]{SCCGowdy}.
\end{remark}
\begin{remark}
  As is clear from the statement, given a generic $\tn{3}$-Gowdy symmetric vacuum solution, the limit $p_{\perp}$ is negative, except for a finite
  number of points. That the condition $p_{\perp}<0$ is advantageous is obvious in view of
  Definition~\ref{def:ndvacidonbbs}. In fact, if $p_{\perp}<0$, then $p_{1}=p_{\perp}$. But then the eigenvector fields corresponding to $p_{2}$ and
  $p_{3}$ are tangential to the symmetry orbits. Since the symmetry orbits are, in addition, abelian Lie groups, it is clear that the eigenvector
  fields corresponding to $p_{2}$ and $p_{3}$ commute. In particular, the condition $\msO^{1}_{23}=0$ is thus automatically satisfied. 
\end{remark}
\begin{remark}
  Given any $\tn{3}$-Gowdy symmetric vacuum solution, the limits of the eigenvalues of $\mK$ exist. Introduce the notation $p_{\perp}$ as above.
  Then, if $p_{\perp}(\vartheta)\in (-1/3,0)$ for some $\vartheta\in\sn{1}$, there is an open subinterval $J$ of $\sn{1}$ such that, on $J$, all
  the conclusions of the theorem up to, and including, (\ref{eq:Xobhform}) hold. This statement follows by combining the proof of the theorem with
  \cite[Proposition~2, pp.~1186--1187]{SCCGowdy}. 
\end{remark}
\begin{remark}
  In (\ref{eq:Xobhform}), the first and last factor on the right hand side depend on $\vartheta$. In general, the same is thus true of the middle
  factor on the right hand side.
\end{remark}
\begin{remark}
  Given a $\tn{3}$-Gowdy symmetric solution, the vectorfield $X_{1}:=\d_{\vartheta}$ is (up to a sign) determined by the following properties: it is
  tangent to the constant-$t$ hypersurfaces; it is orthogonal to the orbits of the $\tn{2}$-group of symmetries; and it satisfies
  $g(X_{1},X_{1})\cdot g(\rograd_{g}t,\rograd_{g}t)=-1$.
\end{remark}
\begin{remark}
  The last observation of the theorem demonstrates that at a true spike, two eigendirections degenerate into one eigendirection, even though the
  limits of the corresponding eigenvalues are distinct. Moreover, this property can be used to characterise the spikes. 
\end{remark}
\begin{proof}
  The proof is to be found in Section~\ref{section:proofGowdy}.
\end{proof}

\subsection{Acknowledgements}
This research was funded by the Swedish Research Council, dnr. 2017-03863.

\section{Proofs of the results in the spatially homogeneous setting}

The purpose of the present section is to prove Theorems~\ref{thm:convergence} and \ref{thm:dataonsingtosolution}. 

\subsection{Equations}
In the Bianchi class A orthogonal stiff fluid setting, there is an expansion normalised formulation of the equations due to Wainwright and Hsu, cf.
\cite[p.~1415]{wah} (see also \cite[Section~2, pp.~414--415]{BianchiIXattr}):
\begin{subequations}\label{eq:Wainwright-Hsu}
\begin{align}
  N_{1}' = & (q-4\Sigma_{+})N_{1},\label{eq:Noevol}\\
  N_{2}' = & (q+2\Sigma_{+}+2\sqrt{3}\Sigma_{-})N_{2},\label{eq:Ntevol}\\
  N_{3}' = & (q+2\Sigma_{+}-2\sqrt{3}\Sigma_{-})N_{3},\label{eq:Nthevol}\\
  \Sigma_{+}' = & -(2-q)\Sigma_{+}-3S_{+},\label{eq:Spevol}\\
  \Sigma_{-}' = & -(2-q)\Sigma_{-}-3S_{-},\label{eq:Smevol}\\
  \Omega' = & 2(q-2)\Omega.\label{eq:Omegaevol}
\end{align}
\end{subequations}%
Here a prime denotes a derivative with respect to a time coordinate $\tau$; the relation between $\tau$ and the proper time $t$ appearing in, e.g.,
(\ref{eq:gdiagonalSH}) is given in Subsection~\ref{ssection:relatingwhsumetric} below; cf. (\ref{eq:thetaisigmaietcdef}). Moreover,
\begin{equation}\label{eq:qdef}
q=2(\Omega+\Sigma_{+}^{2}+\Sigma_{-}^{2}),
\end{equation}
and
\begin{align}
  S_{+} = & \frac{1}{2}[(N_{2}-N_{3})^{2}-N_{1}(2N_{1}-N_{2}-N_{3})],\label{eq:Spdef}\\
  S_{-} = & \frac{\sqrt{3}}{2}(N_{3}-N_{2})(N_{1}-N_{2}-N_{3}).\label{eq:Smdef}
\end{align}
The Hamiltonian constraint can be written
\begin{equation}\label{eq:Hamcon}
  \Omega+\Sigma_{+}^{2}+\Sigma_{-}^{2}+\frac{3}{4}[N_{1}^{2}+N_{2}^{2}+N_{3}^{2}-2(N_{1}N_{2}+N_{2}N_{3}+N_{3}N_{1})]=1.
\end{equation}
By allowing $\Omega=0$, the above set of equations also includes the vacuum setting. 

\subsection{Relating the $N_{i}$'s and the $a_{i}$'s}\label{ssection:relatingwhsumetric}
Deriving the asymptotics of solutions to (\ref{eq:Wainwright-Hsu})--(\ref{eq:Hamcon}) yields information concerning
$(\Omega,\Sigma_{+},\Sigma_{-},N_{1},N_{2},N_{3})$. However, in the end, we wish to derive the asymptotics of the energy density $\rho$ and the
$a_{i}$ appearing in (\ref{eq:gdiagonalSH}). For this reason, it is of interest to relate the variables of the equations with the $a_{i}$'s.
For a detailed discussion of this topic, we refer the interested reader to \cite{BianchiIXattr}. However, let us here make the following
observations. Assume that we have a metric of the form (\ref{eq:gdiagonalSH}), where $\{\xi^{i}\}$ is the dual of a frame $\{e_{i}\}$ satisfying
(\ref{eq:diagonaln}); the construction in \cite[Definition~21.1, p. 489]{BianchiIXattr} is such that this is
automatically satisfied. Next, let $\theta$ denote the mean curvature. In a neighbourhood of a singularity, we can assume $\theta>0$, and we
do so in what follows. Let $E_{i}:=a_{i}^{-1}e_{i}$ (no summation). Then $\d_{t}$, combined with $\{E_{i}\}$, is an orthonormal frame. Moreover,
if $\{i,j,k\}=\{1,2,3\}$,
\[
  [E_{i},E_{j}]=\e_{ijk}o_{k}E_{k}
\]
(no summation on $k$). With this notation,
\begin{equation}\label{eq:okNkdef}
  o_{k}=\frac{a_{k}}{a_{i}a_{j}}n_{k},\ \ \
  N_{k}:=\frac{o_{k}}{\theta}
\end{equation}
(no summation on $k$ in the first equality), assuming $\{i,j,k\}=\{1,2,3\}$; the last equality is the definition of the $N_{k}$ appearing in
(\ref{eq:Wainwright-Hsu}). We also define $\Omega$ by $\Omega=3\rho/\theta^{2}$; cf. the statement of Theorem~\ref{thm:convergence}. Define
$\theta_{i}$, $\sigma_{i}$, $\Sigma_{i}$ and a time coordinate $\tau$ (up to a constant) via
\begin{equation}\label{eq:thetaisigmaietcdef}
  \theta_{i}:=\d_{t}\ln a_{i},\ \ \
  \sigma_{i}:=\theta_{i}-\theta/3,\ \ \
  \Sigma_{i}:=\sigma_{i}/\theta,\ \ \
  dt/d\tau=3/\theta.
\end{equation}
Finally, define $\Sigma_{\pm}$ by
\begin{equation}\label{eq:Sigmapmdef}
  \Sigma_{+}:=3(\Sigma_{2}+\Sigma_{3})/2,\ \ \ \Sigma_{-}:=\sqrt{3}(\Sigma_{2}-\Sigma_{3})/2
\end{equation}
and note that
\begin{equation}\label{eq:thetaprime}
  \d_{\tau}\theta=-(1+q)\theta;
\end{equation}
cf. \cite[(139), p.~487]{BianchiIXattr}, where $q$ is given by (\ref{eq:qdef}). 

Next, we wish to turn things around: starting with information concerning the $N_{i}$, $\Sigma_{\pm}$ and $\Omega$, we wish to draw conclusions
concerning the $a_{i}$ etc. Note, due to (\ref{eq:okNkdef}), that $N_{k}$ is a constant multiple of $a_{k}/(\theta a_{i}a_{j})$. Note also that the
$a_{i}$ can be reconstructed from $\theta$ and the three expressions of the form $a_{k}/(\theta a_{i}a_{j})$. Alternately, if the $n_{k}$ are non-zero,
then the $a_{i}$ can be reconstructed from the $N_{i}$, the $n_{i}$ and $\theta$. However, if there are vanishing $n_{i}$'s, we have to proceed
differently. Let $m_{k}=\ln |n_{k}|$ if $n_{k}\neq 0$ and $m_{k}=0$ otherwise, and define
\begin{equation}\label{eq:mukdef}
  \mu_{k}:=\ln\frac{a_{k}}{\theta a_{i}a_{j}}+m_{k}.
\end{equation}
Then the $a_{i}$ can be reconstructed from the $\mu_{i}$, the $n_{i}$ and $\theta$. It can also be deduced that 
\begin{equation}\label{eq:mukder}
  \d_{\tau}\mu_{1}=q-4\Sigma_{+},\ \ \
  \d_{\tau}\mu_{2}=q+2\Sigma_{+}+2\sqrt{3}\Sigma_{-},\ \ \
  \d_{\tau}\mu_{3}=q+2\Sigma_{+}-2\sqrt{3}\Sigma_{-}.
\end{equation}
Moreover, $N_{k}$ can be written $N_{k}=\e_{k}e^{\mu_{k}}$ (no summation), where $\e_{k}$ equals $-1$, $0$ or $1$. 

\subsection{Asymptotics, spatially homogeneous setting}\label{ssection:asymptoticshs}
The purpose of the present subsection is to prove Theorem~\ref{thm:convergence}.

\begin{proof}[Theorem~\ref{thm:convergence}]
  Let $(G,\bge,\bk,\rho_{0})$ be Bianchi class A orthogonal stiff fluid initial data as in Definition~\ref{def:Bianchiid}. As mentioned in connection
  with Definition~\ref{def:Bianchiid}, there is then a corresponding development such that the metric takes the
  form (\ref{eq:gdiagonalSH}). The statements concerning the existence time and the behaviour of the mean curvature follow from
  \cite[Lemmas~21.5--21.8, pp.~491--493]{BianchiIXattr}, \cite[Lemmas~22.4 and 22.5, pp.~497--498]{BianchiIXattr} and
  (\ref{eq:thetaprime}). 

  Next, note that in the non-vacuum setting, the $N_{i}$ converge to $0$ as $\tau\rightarrow -\infty$ (corresponding to $t\rightarrow 0+$),
  $\Sigma_{\pm}\rightarrow \sigma_{\pm}$ and $\Omega\rightarrow \omega$, where $\omega+\sigma_{+}^{2}+\sigma_{-}^{2}=1$; cf.
  \cite[Theorem~19.1, p.~478]{BianchiIXattr}. In particular the right hand sides of (\ref{eq:mukder}) converge to $2-4\sigma_{+}$,
  $2+2\sigma_{+}+2\sqrt{3}\sigma_{-}$ and $2+2\sigma_{+}-2\sqrt{3}\sigma_{-}$ respectively. Moreover, if $N_{i}\neq 0$, the corresponding expression is
  $>0$; i.e., if $N_{1}\neq 0$, then $2-4\sigma_{+}>0$ etc. This, again, follows from \cite[Theorem~19.1, p.~478]{BianchiIXattr}. The same conclusion
  holds in the vacuum setting, if one excludes the solutions with horizons, quotients of Minkowski space, as well as Bianchi types VIII and IX; cf.
  \cite[Sections~22.6--22.8, pp.~238--241]{RinCauchy} and \cite[Theorem~24.12, p.~258]{RinCauchy}.

  Due to the above observations and (\ref{eq:Wainwright-Hsu}), all the $N_{i}$ converge to zero exponentially.
  Combining this observation with (\ref{eq:qdef}) and (\ref{eq:Hamcon}) yields the conclusion that $q-2$ converges to zero exponentially. Combining
  this observation with (\ref{eq:Wainwright-Hsu}) yields the conclusion that $\Sigma_{\pm}-\sigma_{\pm}$ and $\Omega-\omega$ converge to zero exponentially.
  Before proceeding, it is of interest to relate $t$, $\tau$ and $\theta$. Integrating (\ref{eq:thetaprime}), keeping the fact that $q-2$ decays
  exponentially in mind, yields
  \[
  \theta(\tau)=C_{0}\exp[-3\tau+O(e^{\epsilon\tau})]
  \]
  (in what follows, the exact value of $\epsilon>0$ is not important, and may change from line to line). 
  Combining this equality with (\ref{eq:thetaisigmaietcdef}) and the fact that $\tau\rightarrow-\infty$ corresponds to $t\rightarrow 0+$,
  \begin{equation}\label{eq:toftau}
    t(\tau)=\int_{-\infty}^{\tau}\frac{3}{C_{0}}\exp[3s+O(e^{\epsilon s})]ds=C_{0}^{-1}\exp[3\tau+O(e^{\epsilon\tau})].
  \end{equation}
  Note, in particular, that exponential decay in $\tau$ corresponds to polynomial decay in $t$ and that $t\theta-1$ decays
  exponentially in $\tau$ (and polynomially in $t$). 

  Next, note that with respect to the frame $\{e_{i}\}$ and co-frame $\{\xi^{i}\}$, the expansion normalised Weingarten map can be written
  \[
  \mK_{j}^{\phantom{j}i}=(\theta_{j}/\theta)\delta^{i}_{j}=(\Sigma_{j}+1/3)\delta^{i}_{j}
  \]
  (no summation) using the notation introduced in (\ref{eq:thetaisigmaietcdef}). By the above, there are constants $p_{i}$ such that if we
  let $\msK_{j}^{\phantom{j}i}=p_{j}\delta^{i}_{j}$ (no summation), then $\mK-\msK=O(t^{\epsilon})$. Moreover, $\tr\msK=1$. Next, $\Omega-\omega=O(t^{\epsilon})$
  and the fact that $\omega+\sigma_{+}^{2}+\sigma_{-}^{2}=1$ translates to $\tr\msK^{2}+2\omega/3=1$. Next, note that
  $\d_{\tau}\ln a_{i}=3(\theta_{i}/\theta)=3\Sigma_{i}+1$. This means that there are constants $\alpha_{i}$ such that
  \[
  \ln a_{i}=3p_{i}\tau+\alpha_{i}+O(e^{\epsilon\tau})=p_{i}\ln t+\beta_{i}+O(t^{\epsilon}),
  \]
  where $\beta_{i}=\alpha_{i}+p_{i}\ln C_{0}$, and we appealed to (\ref{eq:toftau}) in the last step. Changing the basis $\{e_{i}\}$ by multiplying each
  of its elements by a constant, we can ensure that (\ref{eq:tmpiailimit}), (\ref{eq:gasymptoticformthm}), (\ref{eq:limitofchhsphom}) as well as
  Remark~\ref{remark:quantitativelimits} hold. That $\msK$ is symmetric with respect to $\bh$ is an immediate consequence of the fact that both are
  diagonal with respect to $\{e_{i}\}$. The fact that $\rodiv_{\bh}\msK=0$ is an immediate consequence of \cite[Lemma~19.13, p.~210]{RinCauchy}, the
  fact that (\ref{eq:diagonaln}) holds and the fact that $\msK$ is diagonal with respect to $\{e_{i}\}$. Finally, the conclusion concerning
  $\msO^{k}_{ij}$ is an immediate consequence of the fact that if $N_{1}\neq 0$, then $2-4\sigma_{+}>0$ etc.; cf. the above comments. 
\end{proof}

\subsection{Data on the singularity and adapted variables}
Our next goal is to prove Theorem~\ref{thm:dataonsingtosolution}. As a first step in that direction, we prove that we can specify data on the singularity
for solutions to (\ref{eq:Wainwright-Hsu})--(\ref{eq:Hamcon}). The data we specify are $(\omega,\sigma_{+},\sigma_{-})\in\rn{3}$,
$(m_{1},m_{2},m_{3})\in\rn{3}$ and $\e_{i}\in \{-1,0,1\}$, $i=1,2,3$. We assume that
\[
\omega+\sigma_{+}^{2}+\sigma_{-}^{2}=1.
\]
Moreover, if $\e_{i}\neq 0$, we assume $p_{i}>0$, where $p_{i}$, $i=1,2,3$, are defined by 
\begin{equation}\label{eq:pisitosigmapm}
  p_{1}=\frac{1}{3}-\frac{2}{3}\sigma_{+},\ \ \
  p_{2}=\frac{1}{3}+\frac{1}{3}\sigma_{+}+\frac{1}{\sqrt{3}}\sigma_{-},\ \ \
  p_{3}=\frac{1}{3}+\frac{1}{3}\sigma_{+}-\frac{1}{\sqrt{3}}\sigma_{-}.
\end{equation}

\begin{prop}\label{prop:variableexandunique}
  Let $(\omega,\sigma_{+},\sigma_{-})$, $(m_{1},m_{2},m_{3})$ and $(\e_{1},\e_{2},\e_{3})$ satisfy the conditions stated above. Then there is a unique
  solution to (\ref{eq:Wainwright-Hsu})--(\ref{eq:Hamcon}) such that $N_{k}\neq 0$ if and only if $\e_{k}\neq 0$, such that $N_{k}$, if non-zero,
  has the same sign as $\e_{k}$, and such that
  \begin{equation}\label{eq:desiredlimit}
    \lim_{\tau\rightarrow-\infty}[\Omega(\tau),\Sigma_{+}(\tau),\Sigma_{-}(\tau),\nu_{1}(\tau),\nu_{2}(\tau),\nu_{3}(\tau)]=(\omega,\sigma_{+},\sigma_{-},0,0,0),
  \end{equation}
  where
  \begin{equation}\label{eq:Niitonui}
    \nu_{i} :=  \mu_{i}-6p_{i}\tau-m_{i}.
  \end{equation}
  Here, if $\e_{k}\neq 0$, $\mu_{k}$ is chosen so that $N_{k} =\e_{k}e^{\mu_{k}}$ (no summation). In particular, $\mu_{k}$ then satisfies the
  corresponding equation in (\ref{eq:mukder}). If $\e_{k}=0$, $\mu_{k}$ is defined to be the solution to the corresponding equation in (\ref{eq:mukder})
  with the property that $\mu_{k}-6p_{k}\tau\rightarrow 0$ as $\tau\rightarrow-\infty$. 
\end{prop}
\begin{remark}
  In case $\e_{k}=0$, it is not obvious that there is a solution $\mu_{k}$ to the corresponding equation in (\ref{eq:mukder}) such that
  $\mu_{k}-6p_{k}\tau\rightarrow 0$ as $\tau\rightarrow-\infty$. However, due to the conclusions (apart from the ones concerning $\mu_{k}$ and $\nu_{k}$)
  and the equations (\ref{eq:Noevol})--(\ref{eq:Nthevol}), it follows that all the $N_{i}$ converge to zero exponentially. This means that $q-2$ converges
  to zero exponentially, and that $\Sigma_{\pm}$ converge exponentially. From this, the desired conclusion can be deduced. 
\end{remark}
\begin{proof}
  We begin by demonstrating that it is sufficient to consider a reduced system.

  \textbf{The reduced system.} Note that if we have a solution to (\ref{eq:Wainwright-Hsu})--(\ref{eq:Hamcon}), then
  \begin{equation}\label{eq:qsecondvers}
    q=2-\frac{3}{2}[N_{1}^{2}+N_{2}^{2}+N_{3}^{2}-2(N_{1}N_{2}+N_{2}N_{3}+N_{3}N_{1})].
  \end{equation}
  In case $\omega>0$, we therefore replace $q$ in (\ref{eq:Noevol})--(\ref{eq:Smevol}) by the right hand side of (\ref{eq:qsecondvers}). The advantage
  of this substitution is that we then can consider the equations (\ref{eq:Noevol})--(\ref{eq:Smevol}) to be autonomous and unconstrained. However,
  once we have constructed a solution to (\ref{eq:Noevol})--(\ref{eq:Smevol}) with the desired asymptotics (for all the variables but $\Omega$), we then
  have to demonstrate that it can be interpreted as a solution to (\ref{eq:Wainwright-Hsu})--(\ref{eq:Hamcon}) with the desired asymptotics. This can be
  achieved as follows. Define $\Omega$ so that the Hamiltonian constraint (\ref{eq:Hamcon}) holds. Then all of
  (\ref{eq:Wainwright-Hsu})--(\ref{eq:Hamcon}) except (\ref{eq:Omegaevol}) hold. Under these circumstances, it can be demonstrated, by a straightforward
  but somewhat lengthy calculation, that (\ref{eq:Omegaevol}) also holds, and we obtain a solution to (\ref{eq:Wainwright-Hsu})--(\ref{eq:Hamcon}). A
  priori, it is not obvious that the $\Omega$ defined in this way is strictly positive. However, due to (\ref{eq:Omegaevol}), $\Omega$ cannot change
  sign in the course of the evolution. If $\Omega\leq 0$ at some point, then $\Omega\leq 0$ for all $\tau$. However, since the Hamiltonian constraint
  (\ref{eq:Hamcon}) holds, since $\Sigma_{+}^{2}+\Sigma_{-}^{2}$ converges to a number $<1$, and since the $N_{i}$ converge to zero, it is clear that
  $\Omega\rightarrow\omega>0$. Thus
  $\Omega(\tau)>0$ for all $\tau$. This leads to the desired existence. If we are able to prove uniqueness of solutions to the subsystem with the
  prescribed asymptotics, we, needless to say, also obtain uniqueness of solutions to the full system. To conclude, it is enough to consider
  (\ref{eq:Noevol})--(\ref{eq:Smevol}) with $q$ given by (\ref{eq:qsecondvers}).

  In case $\omega=0$, we focus on the equations (\ref{eq:Wainwright-Hsu})--(\ref{eq:Hamcon}) with $\Omega=0$. The reason this is justified is that
  for every solution to (\ref{eq:Wainwright-Hsu})--(\ref{eq:Hamcon}) with $\Omega>0$, the limit of $\Omega$ as $\tau\rightarrow-\infty$ exists and
  is strictly positive; cf. \cite[Theorem~19.1, p.~478]{BianchiIXattr}. Such a solution can thus not give rise to the desired asymptotics, and it
  is sufficient to focus on the vacuum case. In the vacuum case, (\ref{eq:Noevol})--(\ref{eq:Smevol}) can already be considered to be autonomous and
  unconstrained, and we do so in what follows. However, we do replace $q$ in (\ref{eq:Noevol})--(\ref{eq:Smevol}) by the right hand side of
  (\ref{eq:qsecondvers}). On the other hand, we then, at the end, have to demonstrate that if we are able to construct solutions to
  (\ref{eq:Noevol})--(\ref{eq:Smevol}) with the desired asymptotics, then they correspond to solutions to (\ref{eq:Wainwright-Hsu})--(\ref{eq:Hamcon})
  with the desired asymptotics. To prove this, let
  \[
  f=1-\Sigma_{+}^{2}-\Sigma_{-}^{2}-\frac{3}{4}[N_{1}^{2}+N_{2}^{2}+N_{3}^{2}-2(N_{1}N_{2}+N_{2}N_{3}+N_{3}N_{1})].
  \]
  Then (\ref{eq:Noevol})--(\ref{eq:Smevol}) can be used to verify that $f'=2(q-2)f$. On the other hand, if the solution to
  (\ref{eq:Noevol})--(\ref{eq:Smevol}) has the desired asymptotics, then all the $N_{i}$ converge to zero exponentially as $\tau\rightarrow-\infty$.
  Since $q$ is given by the right hand side of (\ref{eq:qsecondvers}), this means that $q-2$ is integrable on $(-\infty,0)$. In particular,
  if $f$ is ever non-zero, then $f$ converges to a non-zero number as $\tau\rightarrow-\infty$. On the other hand, if the solution has the
  desired asymptotics, then $f(\tau)\rightarrow 0$ as $\tau\rightarrow-\infty$. To conclude, $f\equiv 0$ and the Hamiltonian constraint
  (\ref{eq:Hamcon}) is satisfied. This means that (\ref{eq:Wainwright-Hsu})--(\ref{eq:Hamcon}) are all satisfied, with $\Omega=0$. This observation
  is sufficient for existence. If we are able to prove uniqueness for the relevant subsystem, we of course also obtain uniqueness for solutions
  to the full system. 

  The above discussion makes it clear that it is sufficient to prove that there is a unique solution to (\ref{eq:Noevol})--(\ref{eq:Smevol}) with
  the desired asymptotics, where $q$ is given by (\ref{eq:qsecondvers}). 

  \textbf{Existence, reduced system.}
  The relation between $\nu_{i}$ and $\mu_{i}$ is given by (\ref{eq:Niitonui}). It will also be convenient to introduce $s_{\pm}:=\Sigma_{\pm}-\sigma_{\pm}$.
  The $\nu_{i}$ satisfy
  \begin{equation}\label{eq:nuieq}
  \nu_{1}'=q-2-4s_{+},\ \ \
  \nu_{2}'=q-2+2s_{+}+2\sqrt{3}s_{-},\ \ \
  \nu_{3}'=q-2+2s_{+}-2\sqrt{3}s_{-}
  \end{equation}
  respectively. Moreover,
  \begin{equation}\label{eq:spmeq}
  s_{+}'=-(2-q)(s_{+}+\sigma_{+})-3S_{+},\ \ \
  s_{-}'=-(2-q)(s_{-}+\sigma_{-})-3S_{-}.
  \end{equation}
  Next, let $\eta_{i}:=6$ if $\e_{i}=0$ and $\eta_{i}:=6p_{i}$ if $\e_{i}\neq 0$. Moreover, let 
  \begin{equation}\label{eq:epsilondef}
    \e:=\min_{i}\{\eta_{i}\}
  \end{equation}
  and let $\tau_{0}\leq 0$. Note that $\e>0$ due to the assumptions. Let $X_{\tau_{0}}$ be the space of continuous functions
  $x_{a}:(-\infty,\tau_{0}]\rightarrow\rn{5}$ such that
  \[
  \sup_{\tau\leq \tau_{0}}\left(\max\left\{e^{-\e\tau}\max_{i}\{|\nu_{i,a}(\tau)|\},e^{-3\e\tau/2}\max_{\pm}\{|s_{\pm,a}(\tau)|\}\right\}\right)\leq 1,
  \]
  where $x_{a}=(\nu_{1,a},\nu_{2,a},\nu_{3,a},s_{+,a},s_{-,a})$. If $x_{a},x_{b}\in X_{\tau_{0}}$, then we define the distance between $x_{a}$ and $x_{b}$ by 
  \begin{equation*}
    \begin{split}
      d(x_{a},x_{b})
      := & \sup_{\tau\leq\tau_{0}}\left(\max\left\{e^{-\e\tau}\max_{i}\{|(\nu_{i,a}-\nu_{i,b})(\tau)|\},
      e^{-3\e\tau/2}\max_{\pm}\{|(s_{\pm,a}-s_{\pm,b})(\tau)|\}\right\}\right),
    \end{split}
  \end{equation*}
  where $x_{a}=(\nu_{1,a},\nu_{2,a},\nu_{3,a},s_{+,a},s_{-,a})$ and $x_{b}=(\nu_{1,b},\nu_{2,b},\nu_{3,b},s_{+,b},s_{-,b})$. Note that $(X_{\tau_{0}},d)$ is a
  complete metric space. 

  Next, we define a map $\Phi$ from $X_{\tau_{0}}$ to itself. Given $x_{a}\in X_{\tau_{0}}$, define $x_{b}$ as follows. First, define $N_{i,a}$
  by (\ref{eq:Niitonui}), where $\nu_{i}$ is replaced by $\nu_{i,a}$ (we here assume that $x_{a}=(\nu_{1,a},\nu_{2,a},\nu_{3,a},s_{+,a},s_{-,a})$).
  In other words, $N_{i,a}=n_{i}e^{\nu_{i,a}+6p_{i}\tau}$, where $n_{i}:=\e_{i}e^{m_{i}}$. Since $q$, $S_{+}$ and $S_{-}$ are polynomials in the $N_{i}$, this
  leads to $q_{a}$ and $S_{\pm,a}$. Next, we define
  \[
  \nu_{1,b}(\tau):=\int_{-\infty}^{\tau}[q_{a}(s)-2-4s_{+,a}(s)]ds
  \]
  and similarly for $\nu_{i,b}$, $i=2,3$; cf. (\ref{eq:nuieq}). Moreover, we define
  \[
  s_{\pm,b}(\tau):=\int_{-\infty}^{\tau}[-(2-q_{a})(s_{\pm,a}+\sigma_{\pm})-3S_{\pm,a}]ds;
  \]
  cf. (\ref{eq:spmeq}). Note that there is a numerical constant $C$ such that
  \[
  |S_{+,a}(\tau)|+|S_{-,a}(\tau)|+|q_{a}-2|\leq C(n_{1}^{2}+n_{2}^{2}+n_{3}^{2})e^{2\e\tau}.
  \]
  Due to this fact, and the fact that $|s_{\pm,a}+\sigma_{\pm}|\leq 2$, it follows that
  \[
  |s_{\pm,b}(\tau)|\leq C\e^{-1}(n_{1}^{2}+n_{2}^{2}+n_{3}^{2})e^{2\e\tau},
  \]
  where $C$ is a numerical constant. In particular,
  \[
  \sup_{\tau\leq\tau_{0}}(e^{-3\e\tau/2}|s_{\pm,b}(\tau)|)\leq C\e^{-1}(n_{1}^{2}+n_{2}^{2}+n_{3}^{2})e^{\e\tau_{0}/2}.
  \]
  Choosing $\tau_{0}$ close enough to $-\infty$, it is clear that the right hand side is $\leq 1$. Next,
  \begin{equation*}
    \begin{split}
      |\nu_{1,b}(\tau)| \leq & \int_{-\infty}^{\tau}[|q_{a}(s)-2|+4|s_{+,a}(s)|]ds
      \leq C\e^{-1}(n_{1}^{2}+n_{2}^{2}+n_{3}^{2})e^{2\e\tau}+\frac{8}{3\e}e^{3\e\tau/2},
    \end{split}
  \end{equation*}
  where we used the fact that $|s_{+,a}(s)|\leq e^{3\e s/2}$. In particular,
  \[
  \sup_{\tau\leq\tau_{0}}[e^{-\e\tau}|\nu_{1,b}(\tau)|]\leq C\e^{-1}(n_{1}^{2}+n_{2}^{2}+n_{3}^{2})e^{\e\tau_{0}}+\frac{8}{3\e}e^{\e\tau_{0}/2}.
  \]
  Choosing $\tau_{0}$ close enough to $-\infty$, the right hand side is $\leq 1$. The arguments concerning $\nu_{i,b}$, $i=2,3$,
  are similar. Letting $x_{b}=(\nu_{1,b},\nu_{2,b},\nu_{3,b},s_{+,b},s_{-,b})$, it follows that $x_{b}\in X_{\tau_{0}}$. 

  Next, we wish to estimate $d[\Phi(x_{a}),\Phi(x_{b})]$. To this end, we need to estimate differences such as
  \[
  N_{i,a}N_{j,a}-N_{i,b}N_{j,b}=n_{i}n_{j}e^{2\e_{ij}\tau}e^{\nu_{i,b}+\nu_{j,b}}(e^{\nu_{i,a}-\nu_{i,b}+\nu_{j,a}-\nu_{j,b}}-1),
  \]
  where $\e_{ij}\geq\e$ can be calculated in terms of the $\sigma_{\pm}$. Since $\tau\leq 0$, we know that $|\nu_{i,a}|\leq 1$ etc. In particular,
  \begin{equation*}
    \begin{split}
      |N_{i,a}N_{j,a}-N_{i,b}N_{j,b}| \leq & |n_{i}n_{j}|e^{2\e_{ij}\tau}e^{2}(e^{|\nu_{i,b}-\nu_{i,a}|+|\nu_{j,b}-\nu_{j,b}|}-1)\\
      \leq & |n_{i}n_{j}|e^{2\e_{ij}\tau}e^{4}(|\nu_{i,b}-\nu_{i,a}|+|\nu_{j,b}-\nu_{j,b}|)\\
      \leq & 2e^{4}|n_{i}n_{j}|e^{(2\e_{ij}+\e)\tau}d(x_{a},x_{b})
    \end{split}    
  \end{equation*}
  for $\tau\leq \tau_{0}$. Thus
  \[
  \int_{-\infty}^{\tau}|N_{i,a}N_{j,a}-N_{i,b}N_{j,b}|ds\leq 2e^{4}(2\e_{ij}+\e)^{-1}|n_{i}n_{j}|e^{(2\e_{ij}+\e)\tau}d(x_{a},x_{b}).
  \]
  Due to arguments of this type,
  \begin{equation}\label{eq:qamqbestetc}
    \begin{split}
      & e^{-3\e\tau/2}\int_{-\infty}^{\tau}(|q_{a}(s)-q_{b}(s)|+|S_{+,a}(s)-S_{+,b}(s)|+|S_{-,a}(s)-S_{-,b}(s)|)ds\\
      \leq & C\e^{-1}(n_{1}^{2}+n_{2}^{2}+n_{3}^{2})e^{3\e\tau_{0}/2}d(x_{a},x_{b})
    \end{split}
  \end{equation}
  for all $\tau\leq \tau_{0}$, where $C$ is a numerical constant. 

  Consider, using the notation $x_{c}=\Phi(x_{a})$ and $x_{d}=\Phi(x_{b})$,
  \begin{equation*}
    \begin{split}
      e^{-\e\tau}|\nu_{1,c}(\tau)-\nu_{1,d}(\tau)| \leq & e^{-\e\tau}\int_{-\infty}^{\tau}[|q_{a}(s)-q_{b}(s)|+4|s_{+,a}-s_{+,b}|]ds\\
      \leq & C\e^{-1}(n_{1}^{2}+n_{2}^{2}+n_{3}^{2})e^{2\e\tau_{0}}d(x_{a},x_{b})+\frac{8}{3\e}e^{\e\tau_{0}/2}d(x_{a},x_{b})
    \end{split}
  \end{equation*}
  for all $\tau\leq \tau_{0}$, where $C$ is a numerical constant. The arguments concerning $\nu_{i,c}-\nu_{i,d}$, $i=2,3$, are similar.
  Next, estimate
  \begin{equation*}
    \begin{split}
      & e^{-3\e\tau/2}|s_{\pm,c}(\tau)-s_{\pm,d}(\tau)|\\
      \leq & e^{-3\e\tau/2}\int_{-\infty}^{\tau}|(q_{a}-q_{b})(s_{\pm,a}+\sigma_{\pm})
      +(q_{b}-2)(s_{\pm,a}-s_{\pm,b})-3(S_{\pm,a}-S_{\pm,b})|ds
    \end{split}
  \end{equation*}
  for $\tau\leq \tau_{0}$. 
  The contributions from the first and last terms inside the absolute value in the integrand can be estimated by appealing to
  (\ref{eq:qamqbestetc}). What remains to be estimated is
  \begin{equation*}
    \begin{split}
      & e^{-3\e\tau/2}\int_{-\infty}^{\tau}|(q_{b}-2)(s_{\pm,a}-s_{\pm,b})|ds\\
      \leq & e^{-3\e\tau/2}\int_{-\infty}^{\tau}C(n_{1}^{2}+n_{2}^{2}+n_{1}^{2})e^{7\e s/2}d(x_{a},x_{b})ds
      \leq C\e^{-1}(n_{1}^{2}+n_{2}^{2}+n_{1}^{2})e^{2\e\tau}d(x_{a},x_{b})
    \end{split}
  \end{equation*}
  for $\tau\leq \tau_{0}$. Summing up,
  \[
  \sup_{\tau\leq\tau_{0}}[e^{-3\e\tau/2}|s_{\pm,c}(\tau)-s_{\pm,d}(\tau)|]\leq C\e^{-1}(n_{1}^{2}+n_{2}^{2}+n_{3}^{2})e^{3\e\tau_{0}/2}d(x_{a},x_{b}),
  \]
  where $C$ is a numerical constant.

  \textbf{Existence.}
  Combining the above estimates, it is clear that for $\tau_{0}$ close enough to $-\infty$, the following estimate holds:
  $d[\Phi(x_{a}),\Phi(x_{b})]\leq d(x_{a},x_{b})/2$.
  Thus $\Phi$ is a contraction on a non-empty and complete metric space. There is thus a unique fixed point. Denote the fixed point by
  $x=(\nu_{1},\nu_{2},\nu_{3},s_{+},s_{-})$. Then, by definition,
  \begin{subequations}\label{eq:integraleq}
  \begin{align}
    \nu_{1}(\tau) = & \int_{-\infty}^{\tau}[q(s)-2-4s_{+}(s)]ds,\\
    \nu_{2}(\tau) = & \int_{-\infty}^{\tau}[q(s)-2+2s_{+}(s)+2\sqrt{3}s_{-}(s)]ds,\\
    \nu_{3}(\tau) = & \int_{-\infty}^{\tau}[q(s)-2+2s_{+}(s)-2\sqrt{3}s_{-}(s)]ds,\\
    s_{\pm}(\tau) = & \int_{-\infty}^{\tau}[-(2-q)(s_{\pm}+\sigma_{\pm})-3S_{\pm}]ds.
  \end{align}  
  \end{subequations}
  Here $N_{i}=n_{i}e^{\nu_{i}+6p_{i}\tau}$. Given the $N_{i}$, the $S_{\pm}$ are defined by (\ref{eq:Spdef})--(\ref{eq:Smdef})
  and $q$ is defined by (\ref{eq:qsecondvers}). We also define $\Sigma_{\pm}$ by $\Sigma_{\pm}:=s_{\pm}+\sigma_{\pm}$. Then, except for the limit
  of $\Omega$, the equality (\ref{eq:desiredlimit}) is satisfied. Since $x$ is continuous and satisfies (\ref{eq:integraleq}), it can, by an
  inductive argument, be verified that it is smooth (in fact, real analytic). It can also be verified that $N_{i}$ and $\Sigma_{\pm}$ satisfy
  (\ref{eq:Noevol})--(\ref{eq:Smevol}). Due to the comments at the beginning of the proof, existence follows. It remains to demonstrate
  uniqueness. 

  \textbf{Uniqueness.} Assume that there is a solution to (\ref{eq:Wainwright-Hsu})--(\ref{eq:Hamcon}) as in the statement of the proposition.
  Then we can define $\nu_{i}$ and $s_{\pm}$ as above. Combining these functions yields an $x:(-\infty,\tau_{0}]\rightarrow\rn{5}$. If we can prove
  that $x\in X_{\tau_{0}}$ for $\tau_{0}$ close enough to $-\infty$, we obtain uniqueness. However, the proof of this statement is similar to the  
  proof of the fact that $\Phi$ maps $X_{\tau_{0}}$ to itself (for $\tau_{0}$ close enough to $-\infty$) and is left to the reader. 
\end{proof}

\subsection{Geometric existence and uniqueness}\label{ssection:proofofgeometricunique}
The goal of this subsection is to prove Theorem~\ref{thm:dataonsingtosolution}.

\begin{proof}[Theorem~\ref{thm:dataonsingtosolution}]
  Consider non-degenerate quiescent Bianchi class A initial data on the singularity for the Einstein-orthogonal stiff fluid equations, say
  $(G,\bh,\msK,\omega)$.
  Assume $\{e_{k}\}$, $p_{i}$ and $n_{i}$ to be chosen so that (\ref{eq:canonicalbasisBianchiA}) holds. Choose $\sigma_{+}$ and $\sigma_{-}$ so that
  (\ref{eq:pisitosigmapm}) holds and so that $\omega+\sigma_{+}^{2}+\sigma_{-}^{2}=1$.

  \textbf{Uniqueness.} Assume that there is a Bianchi class A vacuum or orthogonal stiff fluid development, say $(M,g,\rho)$, with asymptotics
  corresponding to $(G,\bh,\msK,\omega)$. Then $M=H\times I$, where $H$ is a unimodular Lie group and $I$ is an open interval. Moreover,
  \[
  g=-dt\otimes dt+\textstyle{\sum}_{i=1}^{3}a_{i}^{2}(t)\eta^{i}\otimes\eta^{i}
  \]
  where $\{\eta^{i}\}$ is the dual basis of a basis $\{f_{i}\}$ of the Lie algebra $\mathfrak{h}$ of $H$. In addition, $[f_{i},f_{j}]=\e_{ijk}\bn_{k}f_{k}$
  (no summation on $k$), where $\{i,j,k\}=\{1,2,3\}$. Due to the assumptions concerning the asymptotics of $(M,g,\rho)$, the development has a crushing
  singularity. Moreover, by a translation, if necessary, $t$ can be assumed to equal zero at the singularity; i.e., $\theta(t)\rightarrow\infty$ as
  $t\rightarrow 0+$. In particular, we can thus assume the existence of a $t_{0}>0$ such that $\theta(t)>0$ on $(0,t_{0})$. On this interval, the expansion
  normalised Weingarten map associated with $g$ is well defined. Moreover, it is diagonal, and the $f_{k}$ are eigenvectors. Since the Weingarten maps,
  considered as endomorphisms of the tangent space of $H$, converge to $\msK$, we need to have $H=G$ (as manifolds). Moreover, since the eigenvalues of
  $\msK$ are
  distinct, it is clear that each $f_{k}$ has to be a constant multiple of a corresponding $e_{k}$. We can therefore, without loss of generality, assume
  that $f_{k}=e_{k}$ and $\xi^{i}=\eta^{i}$, where $\{\xi^{i}\}$ is the basis dual to $\{e_{i}\}$. Note that this means that $G$ and $H$ have the same Lie
  group structure. 

  Let $\theta_{i}:=\dot{a}_{i}/a_{i}$. Then, since $\mK e_{i}=(\theta_{i}/\theta)e_{i}$ (no summation) and $\msK e_{i}=p_{i}e_{i}$ (no summation), it is
  clear that $\theta_{i}/\theta\rightarrow p_{i}$ as $t\rightarrow 0+$. Define $\Sigma_{i}$ and $\Sigma_{\pm}$ by
  \[
  \Sigma_{i}:=\theta_{i}/\theta-1/3,\ \ \
  \Sigma_{+}:=\frac{3}{2}(\Sigma_{2}+\Sigma_{3})=-\frac{3}{2}\Sigma_{1},\ \ \
  \Sigma_{-}:=\frac{\sqrt{3}}{2}(\Sigma_{2}-\Sigma_{3}).
  \]
  Then $\Sigma_{\pm}\rightarrow\sigma_{\pm}$. Next, define $N_{i}$ by (\ref{eq:okNkdef}). Finally, let $\rho$ denote the energy density and define
  $\Omega$ and $\tau$ by $\Omega=3\rho/\theta^{2}$, $d\tau/dt=\theta/3$. With these choices, $(\Omega, \Sigma_{\pm},N_{i})$ satisfy
  (\ref{eq:Wainwright-Hsu})--(\ref{eq:Hamcon}). Note that there is a translation ambiguity in the definition of $\tau$ which we will address later. However,
  in case the spacetime is not of Bianchi type IX, the range of $\tau$ is $\rn{}$, and if the spacetime is of Bianchi type IX, the range is
  $(-\infty,\tau_{+})$ for some $\tau_{+}\in\rn{}$; cf. \cite[Lemmas~22.4 and 22.5, pp.~497--498]{BianchiIXattr}. 

  Assume the $p_{i}$ to be ordered so that $p_{1}<p_{2}<p_{3}$. If $p_{1}\leq 0$, then $\msO^{1}_{23}=0$ by assumption, so that $n_{1}=0$. This means that
  $N_{1}=0$. Since the $p_{i}$ are distinct and satisfy the Kasner relations, $p_{3}<1$, so that if $p_{1}\leq 0$, then $p_{2}$ has to satisfy $p_{2}>0$.
  This means that we can always assume that $0<p_{2}<p_{3}$, so that $1+\sigma_{+}+\sqrt{3}\sigma_{-}>0$ and $1+\sigma_{+}-\sqrt{3}\sigma_{-}>0$; cf.
  (\ref{eq:pisitosigmapm}). If $p_{1}>0$, we can, additionally, assume that $1-2\sigma_{+}>0$; cf. (\ref{eq:pisitosigmapm}). Next, note that
  $q\rightarrow 2$ since $\Omega\rightarrow\omega$. Combining these observations with (\ref{eq:Noevol})--(\ref{eq:Nthevol}) yields the conclusion that
  all the $N_{i}$ converge to zero exponentially.

  Due to the above observations, it is clear that all the second degree polynomials in the $N_{i}$ appearing in
  (\ref{eq:Wainwright-Hsu})--(\ref{eq:Hamcon}) converge to zero exponentially. Moreover, the same is true of $q-2$.
  This means that
  \begin{equation}\label{eq:thetaasfoftau}
    \theta(\tau)=\exp\left(\int_{\tau}^{\tau_{0}}[1+q(s)]ds\right)\theta(\tau_{0})=C_{0}e^{-3\tau}[1+O(e^{\e\tau})]
  \end{equation}
  for some constants $C_{0},\e>0$, where $\tau_{0}$ is an element of the existence interval of the solution. At this point, we can fix the translation
  ambiguity in the definition of $\tau$ in order to ensure that $\theta(\tau)=e^{-3\tau}[1+O(e^{\e\tau})]$. Since the singularity occurs at $t=0$
  (so that $t$, considered as a function of $\tau$, satisfies $t(-\infty)=0$), we conclude that
  \begin{equation}\label{eq:ttauas}
    t(\tau)=\int_{-\infty}^{\tau}3e^{3s}[1+O(e^{\e s})]ds=e^{3\tau}[1+O(e^{\e\tau})].
  \end{equation}
  In particular,
  \begin{equation}\label{eq:tthetalimit}
    t\theta=1+O(e^{\e\tau}).
  \end{equation}
  Next, define $\mu_{k}$ by (\ref{eq:mukdef}), and compute, using the fact that $t^{-p_{i}}a_{i}\rightarrow 1$ (or, equivalently,
  the fact that $\theta^{p_{i}}a_{i}\rightarrow 1$) and the fact that (\ref{eq:tthetalimit}) holds,
  \[
  \mu_{k}=\ln\frac{t^{p_{k}}}{t^{-1}t^{p_{i}}t^{p_{j}}}+m_{k}+\dots
  =6p_{k}\tau+m_{k}+\dots,
  \]
  where the dots signify terms that tend to zero as $t\rightarrow 0+$ and we appealed to (\ref{eq:ttauas}) in the last step. Define $\nu_{k}$ by
  $\nu_{k}=\mu_{k}-6p_{k}\tau-m_{k}$.
  At this stage, all the data appearing in the statement of Proposition~\ref{prop:variableexandunique} have been determined. In fact,
  $(\omega,\sigma_{+},\sigma_{-})$ are determined by the asymptotic data. Moreover, $m_{k}=0$ if $n_{k}=0$; $m_{k}=\ln |n_{k}|$ if $n_{k}\neq 0$;
  $\e_{k}=1$ if $n_{k}>0$; $\e_{k}=0$ if $n_{k}=0$; and $\e_{k}=-1$ if $n_{k}<0$. Due to Proposition~\ref{prop:variableexandunique}, the corresponding
  solution to (\ref{eq:Wainwright-Hsu})--(\ref{eq:Hamcon}) is uniquely determined by these data. The mean curvature is not, a priori, uniquely
  determined by this. However, it is fixed up to a multiplicative constant, since the deceleration parameter is uniquely fixed. Moreover, we have
  fixed the time coordinate $\tau$ in such a way that $e^{3\tau}\theta(\tau)$ converges to $1$. Combining these observations, leads to the conclusion
  that $\theta$ is uniquely fixed, so that $\rho$ is uniquely fixed. The same is true for the time coordinate $t$, considered as a function of $\tau$.
  Finally, since $\theta$ and $m_{k}$ are fixed, and since $\mu_{k}$ is uniquely determined, we conclude that the quotients $a_{k}/(a_{i}a_{j})$ are
  uniquely determined by appealing to (\ref{eq:mukdef}). From this, we deduce that the functions $a_{k}$ are uniquely determined.

  \textbf{Existence.}
  It remains to prove existence. Fix $\{e_{k}\}$, $p_{i}$, $n_{i}$, $\omega$ and $\sigma_{\pm}$ as at the beginning of the proof. Then
  $\omega+\sigma_{+}^{2}+\sigma_{-}^{2}=1$. Next, let $\e_{i}$ equal zero if $n_{i}=0$ and $\e_{i}=n_{i}/|n_{i}|$ if $n_{i}\neq 0$. Finally, let
  $m_{k}=\ln |n_{k}|$ if $n_{k}\neq 0$ and $m_{k}=0$ otherwise. Note that, by assumption, if $\e_{i}\neq 0$, then $p_{i}>0$. Appealing to
  Proposition~\ref{prop:variableexandunique}, we obtain a unique corresponding solution to (\ref{eq:Wainwright-Hsu})--(\ref{eq:Hamcon})
  such that the statements of Proposition~\ref{prop:variableexandunique} hold. Next, define $\theta$ by the conditions that
  (\ref{eq:thetaprime}) hold and the condition that (\ref{eq:thetaasfoftau}) hold with $C_{0}=1$. Requiring that $t(\tau)\rightarrow 0$ as
  $\tau\rightarrow-\infty$, we conclude that (\ref{eq:ttauas}) holds. Next, we define $\sigma_{i}:=\theta\Sigma_{i}$, where the $\Sigma_{i}$
  can be calculated in terms of the $\Sigma_{\pm}$. Define $\theta_{i}:=\sigma_{i}+\theta/3$ and define $a_{i}$, up to a constant, by
  $\d_{t}\ln a_{i}=\theta_{i}$. Due to Proposition~\ref{prop:variableexandunique}, we know that $\theta_{i}/\theta$ converges to $p_{i}$.
  Moreover, due to the fact that (\ref{eq:Wainwright-Hsu})--(\ref{eq:Hamcon}) holds, and the fact that the $N_{i}$ converge to zero exponentially,
  we know that the convergence is exponential in $\tau$. This means that $\ln a_{i}=3p_{i}\tau+\alpha_{i}+O(e^{\e\tau})$ for some constants
  $\alpha_{i}$ and $\e>0$. We can fix the choice of constants in the $a_{i}$ in such a way that $\ln a_{i}=3p_{i}\tau+O(e^{\e\tau})$.
  This means that $\ln a_{i}=p_{i}\ln t+O(t^{\e})$. At this stage, we can define $g$ by (\ref{eq:gdiagonalSH}) and $\rho$ by
  $\rho=\theta^{2}\Omega/3$. By the above, it is clear that $\theta\rightarrow\infty$ as $t\rightarrow 0+$. Moreover, it is clear that
  conditions (1)--(3) of Theorem~\ref{thm:dataonsingtosolution} are fulfilled. Finally, the Einstein-orthogonal stiff fluid equations are satisfied by
  construction. In order to justify this statement, note that the equations can be written
  \begin{equation}\label{eq:Riccisfeq}
    \mathrm{Ric}=2\rho dt\otimes dt.
  \end{equation}  
  Due to \cite[Lemma~20.1, p.~214]{RinCauchy}, it is sufficient to verify that the diagonal components of this equality is satisfied (in order
  to arrive at this conclusion we use the fact that the metric is diagonal with respect to $\{e_{i}\}$ and the fact that the frame $\{e_{i}\}$
  satisfies the last equality in (\ref{eq:canonicalbasisBianchiA})); note that the $e_{i}$ and $n_{i}$ appearing in \cite[Lemma~20.1, p.~214]{RinCauchy}
  differ from the objects with the same names here. Next, that the $00$-component of (\ref{eq:Riccisfeq}) is satisfied follows from (\ref{eq:thetaprime}).
  That the trace of the $ij$-equations equals zero follows by combining the fact that the $00$-component is satisfied with (\ref{eq:Hamcon}).
  Finally, that the trace free part of the $ij$-equations is satisfied follows from (\ref{eq:Spevol}) and (\ref{eq:Smevol}). 
\end{proof}

\section{$\tn{3}$-Gowdy vacuum spacetimes}\label{section:proofGowdy}

Finally, we prove Theorem~\ref{thm:gowdy}. 

\begin{proof}[Theorem~\ref{thm:gowdy}]
  The existence of the set $\msG$ is guaranteed by \cite[Proposition~3, p.~1190]{SCCGowdy} and \cite[Theorem~2, p.~1190]{SCCGowdy} (in \cite{SCCGowdy},
  this set corresponds to $\msG_{c}$, which is introduced at the top of \cite[p.~1190]{SCCGowdy}). Due to the definition of $\msG$, it follows that,
  excluding a finite number of points $\vartheta_{i}\in\sn{1}$, $i=1,\dots,m$, the function $P_{\tau}(\vartheta,\tau)$ converges to a number belonging
  to $(0,1)$ as $\tau\rightarrow\infty$, where $\tau:=-\ln t$. Due to the arguments presented at the beginning of \cite[Section~C.4.5]{RinWave}, it
  follows that if $\vartheta\in S_{1}$ (where $S_{1}=\sn{1}-\cup_{i=1}^{m}\{\vartheta_{i}\}$), there is an open neighbourhood $I$ of $\vartheta$ and smooth
  functions $v_{a}$, $\phi$, $r$ and $Q_{\infty}$ on $I$, where $\varepsilon<v_{a}<1-\varepsilon$ (for a constant $\varepsilon>0$), a constant $\eta>0$
  and, for each $k\in\nn{}$, a constant $C_{k}$ such that the following estimates hold
  \begin{subequations}\label{eq:asymptoticsPQ}
    \begin{align}
      \|P_{\tau}(\cdot,\tau)-v_{a}\|_{C^{k}(I)}+\|P(\cdot,\tau)-p(\cdot,\tau)\|_{C^{k}(I)} \leq & C_{k}e^{-\eta\tau},\label{eq:Pexp}\\
      \|e^{2p(\cdot,\tau)}Q_{\tau}(\cdot,\tau)-r\|_{C^{k}(I)}+\left\|e^{2p(\cdot,\tau)}[Q(\cdot,\tau)-Q_{\infty}]+r/(2v_{a})\right\|_{C^{k}(I)}
      \leq & C_{k}e^{-\eta\tau},\label{eq:Qexp}
    \end{align}
  \end{subequations}  
  for all $k\in\mathbb{N}$ and $\tau\geq 0$, where $p(\vartheta,\tau):=v_{a}(\vartheta)\tau+\phi(\vartheta)$. Due to this observation, the
  asymptotics can be calculated as in \cite[Section~C.4.5]{RinWave}. Note, to begin with, that the asymptotic expansion normalised Weingarten
  map is given by
  \[
  \msK=\frac{2}{v_{a}^{2}+3}\left(\begin{array}{ccc} (v_{a}^{2}-1)/2 & 0 & 0 \\ 0 & 1-v_{a}+Q_{\infty}r & Q_{\infty}(Q_{\infty}r-2v_{a})\\
    0 & -r & 1+v_{a}-Q_{\infty}r\end{array}\right).
  \]
  This follows from the estimates at the top of \cite[p.~229]{RinWave}. Combining these estimates with \cite[(C.33), p.~228]{RinWave} yields
  (\ref{eq:mKtomsK}). Next, note that the eigenvalues of $\msK$ are given by
  \begin{equation}\label{eq:pidefinition}
    p_{1}=\frac{v_{a}^{2}-1}{v_{a}^{2}+3},\ \ \
    p_{2}=\frac{2(1-v_{a})}{v_{a}^{2}+3},\ \ \
    p_{3}=\frac{2(1+v_{a})}{v_{a}^{2}+3},
  \end{equation}
  so that $p_{1}<p_{2}<p_{3}$, $\tr\msK=1$ and $\tr\msK^{2}=1$. Three corresponding eigenvector fields are given by
  \begin{equation}\label{eq:XAdef}
    X_{1}=\d_{\vartheta},\ \ \
    X_{2}=(Q_{\infty}r-2v_{a})\d_{x}-r\d_{y},\ \ \
    X_{3}=Q_{\infty}\d_{x}-\d_{y}.
  \end{equation}
  The associated dual basis is given by
  \begin{equation}\label{eq:YAdef}
    Y^{1}=d\vartheta,\ \ \
    Y^{2}=-\frac{1}{2v_{a}}(dx+Q_{\infty}dy),\ \ \
    Y^{3}=\frac{1}{2v_{a}}[rdx+(Q_{\infty}r-2v_{a})dy].
  \end{equation}
  Next, we wish to calculate
  \begin{equation}\label{eq:gXoXo}
    \theta^{2p_{1}}g(X_{1},X_{1})=\theta^{2p_{1}}t^{-1/2}e^{\lambda/2}=\theta^{2p_{1}}e^{\tau/2}e^{\lambda/2}.
  \end{equation}
  At this point, it is of interest to note that there is a smooth function $\lambda_{\infty}$ such that
  \begin{equation}\label{eq:lambdaas}
    \|\lambda_{\tau}(\cdot,\tau)+v_{a}^{2}\|_{C^{k}(I)}+\|\lambda(\cdot,\tau)+v_{a}^{2}\tau-\lambda_{\infty}\|_{C^{k}(I)}\leq C_{k}e^{-\eta\tau}
  \end{equation}
  for all $\tau\geq 0$; cf. \cite[Section~C.4.5]{RinWave}. Moreover, there is a smooth positive function $\theta_{\infty}$ on $I$ such that
  \begin{equation}\label{eq:lnthetaasapp}
    \left\|\ln\theta-(v_{a}^{2}+3)\tau/4-\ln\theta_{\infty}\right\|_{C^{k}(I)}\leq C_{k}e^{-\eta\tau}
  \end{equation}
  for all $\tau\geq 0$; cf. \cite[(C.33), p.~228]{RinWave}. Combining the above estimates with \cite[(C.6) and (C.7), p.~224]{RinWave}, it can
  be calculated that
  \begin{equation}\label{eq:thetainfrelation}
    \theta_{\infty}=\frac{1}{4}(3+v_{a}^{2})e^{-\lambda_{\infty}/4}.
  \end{equation}
  Taking the logarithm of the right hand side of (\ref{eq:gXoXo}) and appealing to (\ref{eq:lambdaas}) and (\ref{eq:lnthetaasapp})
  results in the expression
  \[
  2p_{1}\ln\theta+\frac{\tau}{2}+\frac{\lambda}{2}
  =\frac{p_{1}}{2}(v_{a}^{2}+3)\tau+2p_{1}\ln\theta_{\infty}+\frac{\tau}{2}-\frac{v_{a}^{2}}{2}\tau+\frac{\lambda_{\infty}}{2}+O_{k}(e^{-\eta\tau}).
  \]
  Since the terms on the right hand side involving a factor of $\tau$ cancel, we conclude that
  \[
  \left\|\theta^{2p_{1}}g(X_{1},X_{1})-b_{11}\right\|_{C^{k}}\leq C_{k}e^{-\eta\tau}
  \]
  for all $\tau\geq 0$, where
  \[
  b_{11}:=\theta_{\infty}^{2p_{1}}e^{\lambda_{\infty}/2}.
  \]
  Next, consider
  \[
  g(X_{2},X_{2})=te^{P}(Q_{\infty}r-2v_{a})^{2}-2r(Q_{\infty}r-2v_{a})te^{P}Q+r^{2}te^{P}Q^{2}+r^{2}te^{-P}.
  \]
  Note that $Q$ converges exponentially in any $C^{k}$-norm to $Q_{\infty}$ and that $e^{-P}$ converges exponentially to zero. What remains to
  be calculated is thus
  \[
  \ln\left(\theta^{2p_{2}}te^{P}\right)=2p_{2}\ln\theta-\tau+P=2p_{2}\ln\theta_{\infty}+\phi+O_{k}(e^{-\eta\tau}).
  \]
  To conclude
  \[
  \left\|\theta^{2p_{2}}g(X_{2},X_{2})-b_{22}\right\|_{C^{k}}\leq C_{k}e^{-\eta\tau}
  \]
  for all $\tau\geq 0$, where
  \begin{equation}\label{eq:bttdef}
    b_{22}:=\theta_{\infty}^{2p_{2}}e^{\phi}(Q_{\infty}r-2v_{a})^{2}-2r(Q_{\infty}r-2v_{a})\theta_{\infty}^{2p_{2}}e^{\phi}Q_{\infty}
    +r^{2}\theta_{\infty}^{2p_{2}}e^{\phi}Q_{\infty}^{2}=4v_{a}^{2}\theta_{\infty}^{2p_{2}}e^{\phi}.
  \end{equation}
  Next, consider
  \[
  g(X_{2},X_{3})=te^{P}[r(Q-Q_{\infty})^{2}+2v_{a}(Q-Q_{\infty})]+rte^{-P}.
  \]
  To begin with, consider
  \[
  \ln\left(\theta^{2p_{3}}te^{-P}\right)=2p_{3}\ln\theta-\tau-P=2p_{3}\ln\theta_{\infty}-\phi+O_{k}(e^{-\eta\tau}).
  \]
  On the other hand
  \[
  \theta^{2p_{3}}g(X_{2},X_{3})=\theta^{2p_{3}}te^{-P}\left(re^{2P}(Q-Q_{\infty})^{2}+2v_{a}e^{2P}(Q-Q_{\infty})+r\right).
  \]
  Due to (\ref{eq:asymptoticsPQ}), the first term inside the paranthesis on the right hand side converges to zero exponentially in
  any $C^{k}$-norm and the second term converges to $-r$ exponentially in any $C^{k}$-norm. In particular, we conclude that
  \[
  \left\|\theta^{2p_{3}}g(X_{2},X_{3})-b_{23}\right\|_{C^{k}}\leq C_{k}e^{-\eta\tau}
  \]
  for all $\tau\geq 0$, where $b_{23}=0$. Finally, consider
  \[
  g(X_{3},X_{3})=te^{P}(Q-Q_{\infty})^{2}+te^{-P}.
  \]
  By arguments similar to the above, we conclude that
  \[
  \left\|\theta^{2p_{3}}g(X_{3},X_{3})-b_{33}\right\|_{C^{k}}\leq C_{k}e^{-\eta\tau}
  \]
  for all $\tau\geq 0$, where
  \begin{equation}\label{eq:bththdef}
    b_{33} := \theta_{\infty}^{2p_{3}}e^{-\phi}.
  \end{equation}
  At this stage, we can define
  \[
  \bh:=\textstyle{\sum}_{A}b_{AA}Y^{A}\otimes Y^{A}.
  \]
  The above calculations imply that $\bh$ is a smooth Riemannian metric on $S_{1}\times\tn{2}$. Moreover, $\msK$ is symmetric with respect to $\bh$;
  this is an immediate consequence of the fact that $\bh$ is diagonal with respect to $\{X_{A}\}$. At this stage, we can renormalise $\{X_{A}\}$
  in order to obtain a frame $\{\msX_{A}\}$ as in the statement of the theorem. Note that (\ref{eq:chhconvtobh}) holds and that $\msO^{1}_{23}=0$;
  this is an immediate consequence of the fact that $[\msX_{2},\msX_{3}]=0$. That the mean curvature diverges uniformly in the direction of the
  singularity is an immediate consequence of (\ref{eq:lnthetaasapp}).

  It remains to prove that $\rodiv_{\bh}\msK=0$. Denoting the Levi-Civita connection associated with $\bh$ by $\bnabla$,
  \begin{equation*}
    \begin{split}
      (\rodiv_{\bh}\msK)(X_{B}) = & (\bnabla_{X_{A}}\msK)(Y^{A},X_{B})\\
      = & X_{A}(\msK^{A}_{B})-(\bnabla_{X_{A}}Y^{A})(\msK X_{B})-Y^{A}(\msK\bnabla_{X_{A}}X_{B})\\
      = & X_{B}(p_{B})-p_{B}(\bnabla_{X_{A}}Y^{A})(X_{B})-\textstyle{\sum}_{A}p_{A}Y^{A}(\bnabla_{X_{A}}X_{B})
    \end{split}
  \end{equation*}
  (no summation on $B$). Defining $\bGa_{AB}^{C}$ by
  \[
  \bnabla_{X_{A}}X_{B}=\bGa_{AB}^{C}X_{C},
  \]
  this equality can be written
  \[
  (\rodiv_{\bh}\msK)(X_{B})=X_{B}(p_{B})+\textstyle{\sum}_{A}(p_{B}-p_{A})\bGa^{A}_{AB}
  \]
  (no summation on $B$). Defining $m_{A}$ by $\bh(X_{A},X_{A})=e^{2m_{A}}$ (no summation), compute
  \[
  e^{2m_{A}}\bGa_{AB}^{A}=\bh(\bnabla_{X_{A}}X_{B},X_{A})=\bh([X_{A},X_{B}]+\bnabla_{X_{B}}X_{A},X_{A})=[X_{B}(m_{A})+\g_{AB}^{A}]e^{2m_{A}}
  \]
  (no summation on $A$). Thus
  \[
  \bGa_{AB}^{A}=X_{B}(m_{A})+\g_{AB}^{A}
  \]
  (no summation on $A$). Note that the only non-zero commutators are $[X_{A},X_{B}]$ where one of $A$ and $B$ equals $1$ and the other belongs
  to $\{2,3\}$. Note also that for all $A$ and $B$, $[X_{A},X_{B}]$ belongs to $\mathrm{span}\{X_{2},X_{3}\}$. The only non-zero structure constants
  of the form $\g^{A}_{AB}$ (no summation on $A$) are thus $\g^{2}_{21}$ and $\g^{3}_{31}$. Appealing to (\ref{eq:XAdef}) and
  (\ref{eq:YAdef}), it can be verified that
  \[
  \g^{2}_{21}=\frac{1}{2v_{a}}(rQ_{\infty}'-2v_{a}'),\ \ \
  \g^{3}_{31}=-\frac{rQ_{\infty}'}{2v_{a}}.
  \]
  Due to the above observations, it follows that $(\rodiv_{\bh}\msK)(X_{B})=0$ for $B=2,3$. What remains to be calculated is
  \begin{equation*}
    \begin{split}
      (\rodiv_{\bh}\msK)(X_{1}) = & X_{1}(p_{1})+(p_{1}-p_{2})\bGa^{2}_{21}+(p_{1}-p_{3})\bGa^{3}_{31}\\
      = & X_{1}(p_{1})+p_{1}[X_{1}(m_{2}+m_{3})+\g^{2}_{21}+\g^{3}_{31}]\\
      & -p_{2}X_{1}(m_{2})-p_{3}X_{1}(m_{3})-p_{2}\g^{2}_{21}-p_{3}\g^{3}_{31}.
    \end{split}
  \end{equation*}
  Noting that $m_{2}$ and $m_{3}$ can be calculated by means of (\ref{eq:bttdef}) and (\ref{eq:bththdef}) yields
  \begin{equation*}
    \begin{split}
      (\rodiv_{\bh}\msK)(X_{1}) = & X_{1}(p_{1})+p_{1}X_{1}(\ln\theta_{\infty})-\sum_{i=1}^{3}p_{i}X_{1}(p_{i}\ln\theta_{\infty})
      +\frac{1}{2v_{a}}(p_{3}-p_{2})(v_{a}\phi'+rQ_{\infty}')\\
      = & X_{1}(p_{1})-(p_{2}+p_{3})X_{1}(\ln\theta_{\infty})+\frac{2}{v_{a}^{2}+3}(v_{a}\phi'+rQ_{\infty}')\\
      = & \frac{1}{v_{a}^{2}+3}(\lambda_{\infty}'+2v_{a}\phi'+2rQ_{\infty}'),
    \end{split}
  \end{equation*}
  where we appealed to (\ref{eq:pidefinition}) and (\ref{eq:thetainfrelation}); note that (\ref{eq:pidefinition}) implies the sum of the
  $p_{i}$ equals $1$ and that the sum of the $p_{i}^{2}$ equals $1$. To conclude, the statement that $\rodiv_{\bh}\msK=0$ is equivalent to the
  statement that
  \begin{equation}\label{eq:lambdainfder}
    \lambda_{\infty}'=-2v_{a}\phi'-2rQ_{\infty}'.
  \end{equation}
  On the other hand, combining (\ref{eq:lambdatheta}), (\ref{eq:asymptoticsPQ}) and (\ref{eq:lambdaas}) yields exactly this relation. 
  Thus (\ref{eq:lambdainfder}) holds and $\rodiv_{\bh}\msK=0$.

  Combining $g(\d_{\vartheta},\d_{\vartheta})=t^{-1/2}e^{\lambda/2}$ with \cite[(C.7) and (C.8), p.~224]{RinWave} yields
  \[
  t^{2}\theta^{2}g(\d_{\vartheta},\d_{\vartheta})=\frac{1}{16}\rho_{0}^{2}=(1-\ell_{1})^{-2}\rightarrow (1-p_{1})^{-2}.
  \]
  On the other hand,
  \[
  \theta^{2p_{1}}g(\d_{\vartheta},\d_{\vartheta})=\theta^{2p_{1}}g(X_{1},X_{1})\rightarrow \bh(X_{1},X_{1}).
  \]
  Combining these two observations with the fact that $t^{-2}=e^{2\tau}$ yields the conclusion that
  \begin{equation}\label{eq:thetarelatedtotau}
    e^{2\tau}\theta^{-2(p_{2}+p_{3})}\rightarrow (1-p_{1})^{2}\bh(X_{1},X_{1}),
  \end{equation}
  where we used the fact that $p_{1}-1=-p_{2}-p_{3}$. Next, note that the fact that the time coordinate is areal means that
  \begin{equation}\label{eq:arealcondition}
    t^{-2}(\bge_{xx}\bge_{yy}-\bge_{xy}^{2})=1,
  \end{equation}
  where $\bge_{xx}:=g(\d_{x},\d_{x})$, $\bge_{xy}:=g(\d_{x},\d_{y})$ and $\bge_{yy}:=g(\d_{y},\d_{y})$. Next let $E_{i}=X_{i}/|X_{i}|_{\bh}$, $i=1,2,3$,
  and note that
  \[
  \d_{x}=\bh_{x2}E_{2}+\bh_{x3}E_{3},\ \ \
  \d_{y}=\bh_{y2}E_{2}+\bh_{y3}E_{3},
  \]
  where $\bh_{xA}:=\bh(\d_{x},E_{A})$ etc. With this notation, (\ref{eq:arealcondition}) can be written
  \[
  1=e^{2\tau}\theta^{-2(p_{2}+p_{3})}[\theta^{2(p_{2}+p_{3})}\bge(E_{2},E_{2})\bge(E_{3},E_{3})(\bh_{x2}\bh_{y3}-\bh_{x3}\bh_{y2})^{2}+\dots],
  \]
  where the dots signify terms that converge to zero as $t\rightarrow 0+$. Combining this observation with (\ref{eq:thetarelatedtotau}) yields
  the conclusion that
  \begin{equation}\label{eq:arealonsing}
    1=(1-p_{1})^{2}\bh(X_{1},X_{1})(\bh_{x2}\bh_{y3}-\bh_{x3}\bh_{y2})^{2}.
  \end{equation}
  However, the last factor can be rewritten as $\bh_{xx}\bh_{yy}-\bh_{xy}^{2}$. Thus (\ref{eq:arealonsing}) reads
  \[
  1=(1-p_{1})^{2}\bh(X_{1},X_{1})(\bh_{xx}\bh_{yy}-\bh_{xy}^{2}).
  \]
  Integrating the square root of this equality over $\tn{2}$ yields (\ref{eq:Xobhform}). 

  The points excluded by removing the $\vartheta_{i}$ include both so-called true and false spikes; cf. \cite[Sections~C4.6 and C4.7]{RinWave}.
  However, a false spike can be transferred into the type of behaviour we already discussed above by means of an inversion; cf.
  \cite[Section~C4.6]{RinWave}. Moreover, if $(P_{0},Q_{0})$ and $(P_{1},Q_{1})$ are solutions to (\ref{eq:PQ}) related according to
  \cite[(C.36), p.~230]{RinWave}, then
  \[
  te^{P_{1}}(dx+Q_{1}dy)^{2}+te^{-P_{1}}dy^{2}=te^{P_{0}}(dy+Q_{0}dx)^{2}+te^{-P_{0}}dx^{2}.
  \]
  In other words, the effect of an inversion is simply to interchange the coordinates $x$ and $y$ (note that $\lambda$ is unaffected by an
  inversion). This means that the conclusions derived above hold also for the false spikes. We can therefore assume the $\vartheta_{i}$ to only
  represent true spikes. 

  Next, note that for any $\tn{3}$-Gowdy symmetric vacuum solution, and for any $\vartheta\in\sn{1}$, the eigenvalues $\ell_{A}$, $A=1,2,3$,
  of $\mK$ have the property that $\ell_{A}(\vartheta,t)$ converges to, say, $p_{A}(\vartheta)$ as $t\rightarrow 0+$. This follows from the
  discussion in \cite[Section~C.4.2]{RinWave}. Due to \cite[(C.17), p.~226]{RinWave}, it also follows that the limits satisfy the Kasner
  relations. Most of the statements concerning $p_{\perp}$ follow from the definition of $\msG$ and the discussion in \cite[Section~C.4.2]{RinWave}.
  However, the statement concerning the Kasner map remains to be proven. Due to \cite[(18) and (19), p.~2965]{raw} and the discussion in
  \cite[Section~C.4.2]{RinWave}, it follows that at a non-degenerate true spike, say $\vartheta_{i}$, there is a $v_{i}\in (0,1)$ such that
  \begin{equation}\label{eq:pijdef}
    p_{2,i}=2\frac{1+(1+v_{i})}{(v_{i}+1)^{2}+3},\ \ \
    p_{3,i}=\frac{(1+v_{i})^{2}-1}{(v_{i}+1)^{2}+3},\ \ \
    p_{1,i}=2\frac{1-(1+v_{i})}{(v_{i}+1)^{2}+3}.
  \end{equation}
  Moreover, the limit of $p$ (cf. the statement of the theorem) as $\vartheta\rightarrow\vartheta_{i}$ satisfies
  \begin{equation}\label{eq:pijrobasdef}
    p_{2,i,\robas}=2\frac{1+(1-v_{i})}{(v_{i}-1)^{2}+3},\ \ \
    p_{3,i,\robas}=\frac{(1-v_{i})^{2}-1}{(v_{i}-1)^{2}+3},\ \ \
    p_{1,i,\robas}=2\frac{1-(1-v_{i})}{(v_{i}-1)^{2}+3}.
  \end{equation}
  Defining $\Sigma_{i}=p_{i}-1/3$, we can define $\Sigma_{\pm}$ by (\ref{eq:Sigmapmdef}). Let $\Sigma_{\pm,\roin}$ be the $\Sigma_{\pm}$ associated
  with the $p_{i,j}$'s of (\ref{eq:pijdef}) and $\Sigma_{\pm,\roout}$ be the $\Sigma_{\pm}$ associated with the $p_{i,j,\robas}$'s of (\ref{eq:pijrobasdef}).
  Then it can be calculated that
  \[
  \frac{\Sigma_{-,\roout}}{2-\Sigma_{+,\roout}}=\frac{\Sigma_{-,\roin}}{2-\Sigma_{+,\roin}}.
  \]
  To conclude: (\ref{eq:pijdef}) and (\ref{eq:pijrobasdef}) are the future and past endpoints on the Kasner circle of a Bianchi type II solution;
  cf., e.g., \cite[(22.20), p.~239]{RinCauchy}. Moreover $\Sigma_{-,\roout}/\Sigma_{-,\roin}>1$. This proves the statement of the theorem concerning the
  Kasner map.

  The final statements of the theorem are immediate consequences of the discussion in \cite[Section~C4.7]{RinWave}. 
\end{proof}

\end{document}